\documentclass[canadian,reprint, aps, nofootinbib]{article}
\usepackage[T1]{fontenc}
\usepackage[letterpaper]{geometry}
\usepackage{color}
\usepackage{units}
\usepackage{amsmath}
\usepackage{amssymb}
\usepackage{graphicx}
\usepackage[numbers]{natbib}

\usepackage{ulem}

\makeatletter

\newcommand*\LyXThinSpace{\,\hspace{0pt}}
\let\SF@@footnote\footnote
\def\footnote{\ifx\protect\@typeset@protect
    \expandafter\SF@@footnote
  \else
    \expandafter\SF@gobble@opt
  \fi
}
\expandafter\def\csname SF@gobble@opt \endcsname{\@ifnextchar[
  \SF@gobble@twobracket
  \@gobble
}
\edef\SF@gobble@opt{\noexpand\protect
  \expandafter\noexpand\csname SF@gobble@opt \endcsname}
\def\SF@gobble@twobracket[#1]#2{}
\newcommand{\lyxdot}{.}

\newcommand{\lyxaddress}[1]{
	\par {\raggedright #1
	\vspace{1.4em}
	\noindent\par}
}

\usepackage[plainpages=false, colorlinks=true, anchorcolor=blue, linkcolor=blue, citecolor=blue, bookmarks=false]{hyperref}
\date{}

\makeatother

\usepackage{babel}
\begin{document}

\title{\textbf{Dynamical analysis of the covarying coupling constants in
scalar-tensor gravity}}

\author{Rodrigo R. Cuzinatto,$^{1,2}$\thanks{rodrigo.cuzinatto@unifal-mg.edu.br}
Rajendra P. Gupta,$^{1}$\thanks{rgupta4@uottawa.ca} and Pedro J.
Pompeia$^{\ensuremath{1,3}}$\thanks{pompeia@ita.br}}
\maketitle

\lyxaddress{$^{1}$Department of Physics, University of Ottawa, K1N 6N5, Ottawa,
ON, Canada}

\lyxaddress{$^{2}$Instituto de Ciência e Tecnologia, Universidade Federal de
Alfenas, Rodovia José Aurélio Vilela 11999, CEP 37715-400, Poços de
Caldas, MG, Brazil }

\lyxaddress{$^{\ensuremath{3}}$Departamento de Física, Instituto Tecnológico
de Aeronáutica, Praça Mal. Eduardo Gomes 50, CEP 12228-900, São José
dos Campos, SP, Brazil }

\begin{abstract}
A scalar-tensor theory of gravity is considered wherein the gravitational coupling $G$ and the speed of light $c$ are admitted as space-time functions and combine to form the definition of the scalar field $\phi$. The varying $c$ participates in the definition of the variation of the matter part of the action; it is related to the effective stress-energy tensor which is a result of the requirement of symmetry under general coordinate transformations in our gravity model. The effect of the cosmological coupling $\Lambda$ is accommodated within a possible behaviour of $\phi$. We analyze the dynamics of $\phi$ in the phase space, thereby showing the existence of an attractor point for reasonable hypotheses on the potential $V(\phi)$ and no particular assumption on the Hubble function. The phase space analysis is performed both with the linear stability theory and via the more general Lyapunov's method. Either method lead to the conclusion that the condition $\dot{G}/G=\sigma\left(\dot{c}/c\right)$ where $\sigma=3$ must hold for the rest of cosmic evolution after the system gets to the globally asymptotically stable fixed point and the dynamics of $\phi$ ceases. This result realizes our main motivation: To provide a physical foundation for the phenomenological model admitting $\left(G/G_{0}\right)=\left(c/c_{0}\right)^{3}$ used recently to interpret cosmological and astrophysical data. The thus co-varying couplings $G$ and $c$ impact the cosmic evolution after the dynamical system settles to equilibrium. The secondary goal of our work is to investigate how this impact occurs. This is done by constructing the generalized continuity equation in our scalar-tensor model and considering two possible regimes for the varying speed of light\textemdash decreasing $c$ and increasing $c$\textemdash while solving our modified Friedmann equations.  The solutions to the latter equations make room for radiation- and matter-dominated eras that progress to a dark-energy-type of accelerated expansion.
\end{abstract}


\section{Introduction\label{sec:Introduction}}

The possibility of variation of fundamental physical constants has
been long explored as a means of solving some issues in the description
of nature, especially in astrophysics and cosmology. In spite of being
a controversial idea, it has deep connection with scalar-tensor theories,
themselves a popular candidate for viable modified theories of gravity.
This paper explores the possible interweaved variation of the couplings
$G$ and $c$ (while also accommodating the effects of $\Lambda$)
in a Brans-Dicke-like model. Our motivations will be stated and the
construction of our setup will begin during this introductory section,
which starts off with a brief perspective on the subject of varying
fundamental constants. 

The two most studied constants for their potential variation on cosmological
time scales are the gravitational constant $G$ and the fine structure
constant $\alpha$. While $G$ is a dimensionful constant, $\alpha$
is a dimensionless constant. Uzan \citep{Uzan2003,Uzan2011} has critically
reviewed the subject of the variation of fundamental constants and
discussed it extensively. Some authors (Ellis and Uzan \citep{Ellis2005}, Duff \citep{Duff2002,Duff2014}) strongly argue against dimensionful varying fundamental constants. In an attempt to avoid this criticism we normalize the time derivative of the physical coupling by the value of the associated quantity.

Variable fundamental constant theories can be traced back to at least
the late nineteenth century, e.g., \citep{Thomson1883}, and the first
half of the last century, e.g., \citep{Weyl1919,Eddington1934}. However,
they gained importance after Dirac \citep{Dirac1937,Dirac1938} suggested
potential variation of the gravitational constant $G$ based on his
large number hypothesis. Many observational methods were suggested to constrain an eventual variation of $G$ including neutron star masses
and ages \citep{Thorsett1996}; CMB anisotropies, e.g., \citep{Ooba2017};
big-bang nucleosynthesis abundances, e.g., \citep{Alvey2020}; asteroseismology
\citep{Bellinger2019}; lunar laser ranging, e.g., \citep{Hofmann2018};
the evolution of planetary orbits, e.g., \citep{Genova2018}; binary
pulsars, e.g., \citep{Zhu2019}; supernovae type-Ia (SNeIa) luminosity
evolution, e.g., \citep{Wright2018}; and gravitational-wave observations
on binary neutron stars \citep{Vijaykumar2021}. Almost all of them
have resulted in the constraints on $\dot{G}/G$ well below that predicted
by Dirac. There has been significant development on the theoretical side as
well. Building on the work of Jordan \citep{Jordan1959}, Brans and
Dicke \citep{BransDicke1961} developed a scalar-tensor theory of
gravitation wherein $1/G$ was raised to the status of a scalar field
which could vary spatially and temporally. Brans-Dicke (BD) theory
may be seen as one representative of the class of scalar-tensor theories
in which gravitation manifests itself by both the metric tensor and
a scalar field of geometrical nature \citep{Faraoni2004}. A more
general example within this class is the scalar-tensor theory in Lyra
manifold \citep{Sen1971,Cuzinatto2021}. One of the reasons why scalar-tensor
theories of gravity raise so much interest (see \citep{Alonso2003}
and \citep{Faraoni2004} and references therein) is their well know
equivalence with some modified gravity theories \citep{Petrov2020},
cf. e.g., \citep{Capozziello2011,Cuzinatto2016}. Among these possible
modifications of gravity we mention the theories $f\left(R\right)$
\citep{Sotiriou2010,Nojiri2011,Nojiri2017}, $f\left(R,\nabla^{n}R\right)$
\citep{Cuzinatto2019PRD}, $f\left(R,\Box^{n}R\right)$ \citep{Wands1994},
$f\left(T\right)$ \citep{Ferraro2018,Pompeia2021}, which have been
proposed as attempts to deal with the GR quantization problem \citep{Donoghue2019,Buchbinder2021},
to realize inflationary models \citep{Rodrigues2022},
to address the dark energy problem \citep{Cuzinatto2015}. Refs. \citep{Ballardini2020,Ballardini2021,Braglia2021}
studied the phenomenology of extended Jordar-Brans-Dicke theories
and their implication for cosmological datasets. Moreover, significant astrophysical consequences can be found in the literature within the realm of modified theories of gravity including those in Refs. \cite{Capozziello2002,Yousaf2022,DeFelice2010,Clifton2012}.

The \textit{constancy} of the speed of light is the foundation of special
and general relativity theories, and arguably its variation is the
most contentious issue in physics. However, even Einstein considered
its possible \textit{variation} \citep{Einstein1907}. There are several theories of the variation of the speed light, e.g., those by Dicke \citep{Dicke1957}, Petit \citep{Petit1988}, and Moffatt
\citep{Moffat1993,Moffat1993FoundPhys}. Some of these proposals break Lorentz invariance, e.g., \citep{AlbrechtMagueijo1999,Barrow1999,Moffat2016}, others produce locally invariant theories \citep{Avelino1999,Avelino2000}.

Attempts have also been made to consider the simultaneous variation
of two or more constants, e.g., Ref.~\citep{Maharaj1993} considered varying $G$ and $\Lambda$; variation of $G/c^{2}$ was studied in \citep{Belinchon2003}; $G$ and $\alpha$ were the changing concomitantly in Ref.~\citep{Chakrabarti2022}; co-varying $c$, $G$ and $\Lambda$ were the subject of Refs.~\citep{Franzmann2017,Costa2019,Gupta2019}; the set $\{c,G,\alpha\}$ was allowed to vary in \citep{Eaves2021}; and, finally, Refs.~\citep{Gupta2022SNeIa,Gupta2022VCC} suggested that all the couplings $G$, $c$, $\Lambda$, $\hbar$, and $k_{B}$ must co-vary.

After this contextualization on the subject of varying physical constants,
let us concentrate on the topic to be explored here. Our focus in
this paper is on the variation of $c$ and $G$ and the interrelationship
of their variation. We will see that this interrelationship yields
naturally to an effective cosmological constant. Gupta \citep{Gupta2020Cosmology}
introduced the general ansatz 
\begin{equation}
\frac{\dot{G}}{G}=\sigma\frac{\dot{c}}{c}\label{eq:GuptasAnsatz(G,c)}
\end{equation}
inspired by the general constraint appearing in the proposal by Costa
et al. \citep{Costa2019}. The quantity $\sigma$ is a constant parameter.
The value $\sigma=3$ is strongly favored by several phenomenological
applications of Eq. (\ref{eq:GuptasAnsatz(G,c)}) in cosmology \citep{Gupta2022SNeIa,Gupta2020Cosmology,Gupta2021BBN,Gupta2022Quasars}
and astrophysics \citep{Gupta2022VCC,Gupta2021Lensing,Gupta2021Orbital,Gupta2022FaintSun}.
This preferred value for $\sigma$ is corroborated by other authors,
cf. e.g., Ref. \citep{Eaves2021}. For this reason, we set off to
investigate the fundamental reasons that might be underlying this
fact. 

We do that by assuming a modification in the standard theory of gravity:
General Relativity (GR). This is justified from the fact that GR does
not allow for varying physical constants. In fact, the couplings $G,$
$c$, and $\Lambda$ showing up in Einstein field equations
\begin{equation}
G_{\mu\nu}+\Lambda g_{\mu\nu}=\frac{8\pi G}{c^{4}}T_{\mu\nu}\label{eq:EFE}
\end{equation}
do not vary in time or space. Recall the definition of the Einstein
tensor: $G_{\mu\nu}=R_{\mu\nu}-\frac{1}{2}g_{\mu\nu}R$, where $R_{\mu\nu}$
is the Ricci tensor and $R$ is the curvature scalar. We follow the
metric signature and Riemann tensor definition of Ref. \citep{Carroll2019},
i.e. the Minkowski metric in Cartesian coordinates reads $\left(\eta_{\mu\nu}\right)=\text{diag}\left(-1,+1,+1,+1\right)$,
$R_{\mu\nu}=R_{\hphantom{\rho}\mu\rho\nu}^{\rho}$ with the curvature
tensor calculated by $R_{\hphantom{\rho}\sigma\mu\nu}^{\rho}=\partial_{\mu}\Gamma_{\hphantom{\rho}\nu\sigma}^{\rho}+\Gamma_{\hphantom{\rho}\mu\lambda}^{\rho}\Gamma_{\hphantom{\rho}\nu\sigma}^{\lambda}-\left(\mu\leftrightarrow\nu\right)$,
$R=g^{\mu\nu}R_{\mu\nu}$, and $\Gamma_{\hphantom{\rho}\mu\nu}^{\rho}=\frac{1}{2}g^{\rho\sigma}\left(\partial_{\nu}g_{\sigma\mu}+\partial_{\mu}g_{\nu\sigma}-\partial_{\sigma}g_{\mu\nu}\right)$
stands for the Christoffel symbols built from the metric tensor $g_{\mu\nu}$
and its first-order derivatives. Covariant derivatives are built from
the Christoffel, e.g., $\nabla_{\mu}V^{\nu}=\partial_{\mu}V^{\nu}+\Gamma_{\hphantom{\nu}\mu\rho}^{\nu}V^{\rho}$
for an arbitrary contravariant vector $V^{\nu}$.

Our generalization of GR stems from an action integral that is inspired
in Einstein-Hilbert action,
\begin{equation}
S_{g}=\frac{1}{16\pi}\int d^{4}x\sqrt{-g}\left[\frac{c^{3}}{G}\left(R-2\Lambda\right)\right],\label{eq:Einstein-Hilbert}
\end{equation}
but with the couplings $\left\{ G,c,\Lambda\right\} $ being spacetime
functions. At this point, we emphasize the natural appearance of the
couplings $\left\{ G,c,\Lambda\right\} $ explicitly in the action
integral that is supposed to describe gravity. We feel the urge to
underscore this fact for two reasons. The first is: in the ordinary
approach of GR, both $G$ and $c$ are only multiplicative constants
of the kernel $\left(R-2\Lambda\right)$ and, as such, can be taken
outside the integral with impunity. In fact, people even go further
to use units where $c=1$, the so called natural units. These practices
are very dangerous here because both $G$ and $c$ are space-time
dependent functions; accordingly, they are directly affected by the
variation process soon to be carried out. Second, the fact that (\ref{eq:Einstein-Hilbert})
displays the set $\left\{ G,c,\Lambda\right\} $ justifies why we
consider only these three couplings, instead of bringing about other
fundamental constants as well, such as the Planck constant $\hbar$
and the Boltzmann constant $k_{B}$. Others works considered this
enlarged set from a phenomenological perspective\textemdash see e.g.,
Refs. \citep{Gupta2022SNeIa,Gupta2022VCC,Gupta2020Cosmology,Gupta2021BBN,Gupta2022Quasars,Gupta2022FaintSun,Lee2021}. 

As usual, the four-volume in the action $S_{g}$ is $d^{4}x=dx^{0}d^{3}\mathbf{x}$.
It pays out to stress that the time coordinate $x^{0}=ct$ encompasses
the speed of light $c$, which is allowed to vary in our context.
Therefore, it is paramount to work with $x^{0}$ throughout. In this
way, covariance is guaranteed; caveats like having to deal the opened
$x^{0}$-differential $dx^{0}=d\left(ct\right)=cdt+tdc$ are avoided;
and the speed of light will not appear in the metric components $g_{\mu\nu}$
explicitly. Consequently, the determinant of the metric tensor $g=g\left(x^{\mu}\right)$
will depend explicitly on $x^{0}$, but not on $c$. The time component
$x^{0}$ hides $c$ in a convenient form: if $c$ changes, then a
time reparameterization $t\rightarrow t^{\prime}$ compensates for
this fact in each reference frame. This strategy is somewhat different
from the procedure of the $c$-flation framework developed in Ref.
\citep{Costa2019}. Here, $x^{0}$ has dimensions of length, which
explains the power cube for the speed of light in (\ref{eq:Einstein-Hilbert}):
otherwise the action would not bear the correct dimension of angular
momentum (energy $\times$ time). 

Gupta's proposal in Eq. (\ref{eq:GuptasAnsatz(G,c)}) indicates that
the couplings $G$ and $c$ are intertwined. In fact, it could be
recast as:
\begin{equation}
\frac{\dot{\phi}}{\phi}=0,\label{eq:GuptasAnsatz(phi)}
\end{equation}
where
\begin{equation}
\phi\equiv\frac{c^{3}}{G}\label{eq:phi}
\end{equation}
is a scalar field comprising the gravitational coupling $G$ and the
speed of light $c$. The dot on top of the quantity means a derivative
with respect to $x^{0}$, for instance $\dot{\phi}=\frac{d\phi}{dx^{0}}$. 

The simplest way to satisfy (\ref{eq:GuptasAnsatz(G,c)}) is to admit
that $\dot{\phi}=-\phi\left[\frac{\dot{G}}{G}-3\frac{\dot{c}}{c}\right]=0$
due to constant couplings, in which case $\dot{G}=0$ and $\dot{c}=0$.
This is the way of GR, but it is not the most general possibility
and not the one we will be adopting here. An earlier attempt to relax
the constancy of fundamental couplings in the context of a theory
for gravity was the Brans-Dicke (BD) scalar-tensor theory \citep{BransDicke1961}.
Aiming to fully realize the Machian principle in a geometrical theory
of gravitation, Brans and Dicke proposed a varying gravitational coupling
through the scalar field \citep{Faraoni2004}
\begin{equation}
\phi_{\text{BD}}=\frac{1}{G}.\label{eq:phiBD}
\end{equation}

Our proposal is to extend the scope of BD field to include the spacetime
coupling $c$ as part of its very definition. The presence of $c$
changes the nature of the scalar field $\phi$ as previously envisioned
by Brans and Dicke.\footnote{In fact, we are geometrizing not only the gravitational coupling\textemdash through $G$\textemdash but also the causality coupling $c$.} This is true even from a dimensional point of view. In the Brans-Dicke picture, $\phi_{\text{BD}}$ is measured in units of $\left(\text{mass}\right)\times\left(\text{time}\right)^{2}/\left(\text{length}\right)^{3}$.
In our scalar-tensor gravity, $\phi$ has dimensions of $\left(\text{mass}\right)/\left(\text{time}\right)$.

The fact that a varying $c$ enters the scalar field $\phi$ in Eq.
(\ref{eq:phi}) may be a cause of concern to some. A common criticism
upon varying speed of light proposals is that the Maxwell electrodynamics
should not be tempered with. This is a wise thought, however one that
can be bypassed by arguments such as that by Ellis and Uzan in Ref.
\citep{Ellis2005}. There could be different types of speed of light,
viz. the spacetime speed of light $c_{\text{ST}}$ related to Lorentz
invariance and causality and the electromagnetic speed of light $c_{\text{EM}}$
built from the couplings in electromagnetism: the (vacuum) electric
permittivity $\epsilon_{0}$ and the magnetic permeability $\mu_{0}$.
Here we interpret $c$ as the causality coupling $c_{\text{ST}}$
and ensure that Lorentz invariance (of Maxwell's theory) is not broken
locally by convenient reparameterizations of $t$ at each time slice
of the space-time manifold. Moreover, our approach deals with cosmology
which makes room for a variation of $c$ with respect to the cosmic
time in such a way that $c_{{\rm ST}}=c_{{\rm EM}}=c_{0}$ nowadays.

The main goal of this paper is to investigate if the dynamics of $\phi$
can naturally lead $\dot{G}/G=3\left(\dot{c}/c\right)$ as an attractor
solution. This would give the fundamental reason for why the value
$\sigma=3$ is preferred in several apparently different contexts,
such as cosmology \citep{Gupta2020Cosmology}, solar astrophysics
\citep{Gupta2022FaintSun}, and solar system kinematics \citep{Gupta2021Orbital}.
After the eventual relation between $G\left(x^{0}\right)$ and $c\left(x^{0}\right)$
is established, a natural question poses itself: ``What are the consequences
of $G=G\left(c\right)$ for the cosmic evolution?''. Our secondary
goal in this paper is to answer this question by investigating cosmological
solutions attributable to different epochs in the Universe's thermal
history (e.g., radiation dominated era, dust-matter era, and recent
accelerated regime). We will build our scalar-tensor model in a metric
compatible, four-dimensional, torsion-free, Riemannian manifold. Approaches
considering richer geometrical structure \citep{Cuzinatto2021}, higher-order
derivatives of curvature-based objects \citep{Cuzinatto2016,Cuzinatto2019PRD}
or torsion-based invariants \citep{Pompeia2021} are possible, but
these cases shall be explored elsewhere.

The remainder of the paper is organized as follows. In Section \ref{sec:FieldEquations}
we state the action of our scalar-tensor model, derive the equations
of motion for the tensor field $g_{\mu\nu}$ and the scalar field
$\phi$, which together build the gravitational field. The field equations
are specified for the FLRW metric \citep{Weinberg1972}, since our
main interest will be to study the dynamics of $\phi$ in the cosmological
context. This study is carried out in Section \ref{sec:Dynamics},
where it is shown that an attractor solution leading to $\sigma=3$
is attainable under reasonable hypotheses for the potential $V\left(\phi\right)$
and the dominant matter-energy content during the evolution of $\phi$. \footnote{Both the linear stability theory and Lyapunov's method are briefly reviewed in Subsection \ref{subsec:Stability-theory} for the sake of completeness of our presentation and convenience for the reader.}
This finding comes from the linear stability analysis of our dynamical
system\textemdash performed in Subsection \ref{subsec:Phase-space-Linear-Theory};
the result is confirmed and strengthened in Subsection \ref{subsec:Phase-space-Lyaponov-Method}
where we use the general Lyapunov's method to show that the critical
point $\phi_{\text{eq}}$ leading to $\sigma=3$ is a globally asymptotically
stable fixed point. Subsection \ref{subsec:Discussion} gives some details about
the system's attractor point and emphasizes the consequences for our
modified gravity. The cosmic evolution after $\phi$ reaches equilibrium
is analyzed in Section \ref{sec:Gravity-after-equilibrium} wherein
it is shown that our scalar-tensor model accommodates radiation- and
matter-dominated eras evolving naturally to a de Sitter-type accelerated
expansion, possibly recognized as dark energy. This is true for the
two parameterizations we adopted for the function $c=c\left(x^{0}\right)$,
the first assuming a speed of light that decreases as the universe
expands while the second admits $c$ increasing with the scale factor.
Our final comments appear in Section \ref{sec:Final-comments}. 

\section{Modified gravity with $\phi=c^{3}/G$\label{sec:FieldEquations}}

The motivations in the previous section lead us to describe gravity
from a precisely specified form for the action integral. In fact,
this form is hinted by the coefficient of the curvature scalar $R$
in Eq. (\ref{eq:S}), namely $\left(c^{3}/G\right)$, which was defined
as the scalar field $\phi$. Accordingly, we introduce
\begin{align}
S & =\frac{1}{16\pi}\int d^{4}x\sqrt{-g}\left(\phi R-\frac{\omega}{\phi}\nabla_{\mu}\phi\nabla^{\mu}\phi-V\left(\phi\right)\right)\nonumber \\
 & \hphantom{=}+\int d^{4}x\sqrt{-g}\left(\frac{1}{c}\mathcal{L}_{m}\right),\label{eq:S}
\end{align}
where $\omega$ is a dimensionless constant of order one\footnote{It is argued that Brans-Dicke theory recovers general relativity in the limit $\omega \rightarrow \infty$, cf. e.g. Ref. \cite{Weinberg1972}. However, this is not always the case: Ref. \cite{Romero1993} presents a counter-example.} and $\phi$
is a scalar field. The constant $\omega$ appears as the coefficient
of the kinetic term for the field $\phi$; the denominator of this
term includes $\phi$ for dimensionality consistency. The cosmological
coupling $\Lambda$ was not written explicitly in (\ref{eq:S}) because
it can be included in the potential $V\left(\phi\right)$. In fact,
we will show that a constant $\Lambda$ can be realized in our framework.
The reader can easily check that all terms in (\ref{eq:S}) were crafted
in order to give $S$ the correct dimension of $\left(\text{energy}\right)\times\left(\text{time}\right)$.
This includes inserting the factor $\left(1/c\right)$ explicitly inside
the matter term integral containing the matter Lagrangian density
$\mathcal{L}_{m}$, itself carrying dimensions of energy density.

The action (\ref{eq:S}) is formally the same as Brans-Dicke action.
While Brans-Dicke scalar field is usually interpreted as an effective
$G^{-1}$ (see Ref. \citep{Faraoni2004}), our $\phi$ includes both
$G$ and $c$. Notice, moreover, that (\ref{eq:S}) is not the same
as the actions in the Varying Speed of Light (VSL) proposals\textemdash such
as those in Refs. \citep{Moffat1993,AlbrechtMagueijo1999,Barrow1999}\textemdash for
reasons ranging from different modelling to the basic fact that variations
of the gravitational coupling are disregarded therein. Additionally,
it should be pointed out that (\ref{eq:S}) does not correspond to
the same treatment given in Ref. \citep{Costa2019}; indeed, the covariant
$c$-flation framework considers $G$ and $c$ as independent fields
while here these varying couplings enter the form of a single scalar
field; on top of that, the very form of the action integral is different
both in its gravitational part and in the kinetic term for the scalar
field(s).

Variation of the action in Eq. (\ref{eq:S}) with respect to the metric
$g^{\mu\nu}$ yields the BD-like Field Equations for $g_{\mu\nu}$:\footnote{It is matter of interpretation to label the model explored in this
paper as a Brans-Dicke theory. Our model is not Brans-Dicke if one
considers as Brans-Dicke a scalar-tensor theory where the scalar field
$\phi$ is related strictly to $1/G$. Our model could be said to
be Brans-Dicke if the latter is a scalar-tensor theory for $\phi$
regardless of its interpretations in terms of the fundamental couplings.
We prefer the first interpretation. However, even in the second interpretation,
our paper shows that new physics arises from defining $\phi$ in terms
of both $G$ and $c$: this novel perspective allows for the entangled
variation of $G$ and $c$ in a theoretical structure accommodating
a consistent variational principle, the covariant conservation of
an effective energy momentum tensor, and a fundamental explanation
for recent enticing phenomenological results pointing towards $\dot{G}/G=3\left(\dot{c}/c\right)$.} 

\begin{align}
R_{\mu\nu}-\frac{1}{2}g_{\mu\nu}R+\frac{V}{2\phi}g_{\mu\nu} & =\frac{1}{c}\frac{8\pi}{\phi}T_{\mu\nu}\nonumber \\
 & +\frac{\omega}{\phi^{2}}\left(\nabla_{\mu}\phi\nabla_{\nu}\phi-\frac{1}{2}g_{\mu\nu}\nabla^{\rho}\phi\nabla_{\rho}\phi\right)\nonumber \\
 & +\frac{1}{\phi}\left(\nabla_{\mu}\nabla_{\nu}\phi-g_{\mu\nu}\square\phi\right),\label{eq:BD-likeFE}
\end{align}
where
\begin{equation}
T_{\mu\nu}=\frac{-2}{\sqrt{-g}}\frac{\delta\left(\sqrt{-g}\mathcal{L}_{m}\right)}{\delta g^{\mu\nu}}\label{eq:Tmunu}
\end{equation}
is the stress-energy tensor of ordinary matter. Notice that $T_{\mu\nu}$
in (\ref{eq:Tmunu}) has the regular dimensions of energy density.
However, it should be underscored that the variation of the matter
action
\begin{equation}
S_{m}=\int d^{4}x\sqrt{-g}\left(\frac{1}{c}\mathcal{L}_{m}\right)\label{eq:Sm}
\end{equation}
with respect to $g^{\mu\nu}$ is written as
\begin{equation}
\delta S_{m}=-\int d^{4}x\sqrt{-g}\left(\frac{1}{c}\right)\left(\frac{1}{2}T_{\mu\nu}\right)\delta g^{\mu\nu}\label{eq:deltaSm}
\end{equation}
in the context of our scalar-tensor model. Here the factor $\left(1/c\right)$
must participate the form of $\delta S_{m}$ within the integral sign.
This is not the case in non-varying speed of light scenarios (including
the pure BD theory). The factor $\left(1/c\right)$ accompanying $\mathcal{L}_{m}$
(or $T_{\mu\nu}$) brings about one of the interesting feature in
our scalar-tensor model: large-valued varying-$c$ setups suppress
the ordinary matter contribution to the action; in which case, the
dynamics is mostly determined by $\phi$ through its kinetic and potential
terms. In Eq. (\ref{eq:Tmunu}), no dependence of $\mathcal{L}_{m}$
on the field $\phi$ is assumed. However, it should be noted that
we can not evade a non-trivial coupling with the (varying) speed of
light within the functional derivative entering the definition of
$T_{\mu\nu}$. This feature will be dealt with later---see Subsection \ref{subsec:Tmunu(eff)}; it is related to the definition of an effective stress-energy tensor including $c$. According to Noether's theorem \cite{Noether1918} to every symmetry corresponds a conserved current; in theories of gravity, the conserved current is the own stress-energy tensor and the symmetry is the invariance under general coordinate transformations (realized in practice by infinitesimal translations involving Killing vectors \cite{Aldrovandi2016,Aldrovandi2007}). In the paper
\citep{Costa2019} this delicate point surrounding the underlying symmetry and the associated energy-momentum tensor appears in a different form:
the covariant $c$-flation approach introduces $T_{\mu\nu}$ directly
in the action via a Lagrange multiplier.

Variation of $S$ with respect to the scalar field $\phi$ leads to
the Equation of Motion (EOM) for the scalar part of our scalar-tensor
model:
\begin{equation}
\frac{2\omega}{\phi}\square\phi+R-\frac{\omega}{\phi^{2}}\nabla^{\rho}\phi\nabla_{\rho}\phi-\frac{dV}{d\phi}-16\pi\frac{1}{c^{2}}\frac{dc}{d\phi}\mathcal{L}_{m}=0.\label{eq:EOM-phi}
\end{equation}
This EOM tells that ``$\phi$ is a dynamical field; it changes in
space and time.'' As a consequence of the definition of $\phi$ in
terms of $G$ and $c$, the latter conclusion yields the co-varying
character of the gravitational coupling and the causality coupling.
We will show bellow that the simultaneous variation of $G$ and $c$
occurs in such a way that the condition $\dot{G}/G=3\left(\dot{c}/c\right)$
is satisfied after the system evolves to the equilibrium stable point
in the phase space.

In the following we specify our BD-like theory (in the Jordan frame)
for cosmology\textemdash see e.g., \citep{Faraoni2004}. The main
hypothesis is that $\phi$ depends only on the cosmic ``time'' $x^{0}=ct$
due to the requirements of homogeneity and isotropy. These requirements
demand the FLRW line element:
\begin{equation}
ds^{2}=-\left(dx^{0}\right)^{2}+a^{2}\left(x^{0}\right)\left[\frac{dr^{2}}{1-kr^{2}}+r^{2}\left(d\theta^{2}+\sin^{2}\theta d\varphi^{2}\right)\right].\label{eq:FLRW}
\end{equation}
In this set of coordinates the scale factor $a\left(x^{0}\right)$
bears the dimension of length. Conversely, the comoving coordinates
$\left\{ r,\theta,\varphi\right\} $ are dimensionless quantities.
The curvature of the space section is determined by the parameter
$k=-1,0,+1$ as usual \citep{Weinberg1972}. The Hubble function $H$
is defined as
\begin{equation}
H=\frac{\dot{a}}{a}\label{eq:H}
\end{equation}
and has the dimension of inverse length.

The matter content appearing in both field equations (\ref{eq:BD-likeFE})
and (\ref{eq:EOM-phi}) shall be modelled by the perfect fluid stress-energy
tensor,
\begin{equation}
T_{\hphantom{\mu}\nu}^{\mu}=\text{diag}\left\{ -\varepsilon,p,p,p\right\} ,
\label{eq:Tmunu-perfectfluid}
\end{equation}
where $p$ is the pressure and $\varepsilon$ is the energy density
(both with dimensions of energy per unit volume). 

Due to (\ref{eq:Tmunu-perfectfluid}), the $00$-component of the
$g_{\mu\nu}$ field equations (\ref{eq:BD-likeFE}) reads:\footnote{Additional details on how to calculate Eqs.~(\ref{eq:1stFriedmannEq}), (\ref{eq:Hdot(epsilon,p,phi)}), and (\ref{eq:EOM-phi-Cosmology}) are given in the Appendix.}
\begin{equation}
H^{2}=\frac{1}{c}\frac{8\pi}{3\phi}\varepsilon+\frac{\omega}{6}\left(\frac{\dot{\phi}}{\phi}\right)^{2}-H\frac{\dot{\phi}}{\phi}-\frac{k}{a^{2}}+\frac{V}{6\phi}.
\label{eq:1stFriedmannEq}
\end{equation}
This is the (first) Friedmann equation of our scalar-tensor cosmology
with covarying $G$ and $c$ through the field $\phi$. 

The second Friedmann equation (or acceleration equation) is calculated
by taking the trace of Eq. (\ref{eq:BD-likeFE}), using the curvature
scalar built from the metric in (\ref{eq:FLRW}), utilizing the trace
of the stress energy tensor, and the Eqs. (\ref{eq:EOM-phi}) and
(\ref{eq:1stFriedmannEq}). It reads:
\begin{align}
\dot{H} & =-\frac{1}{c}\frac{8\pi}{\left(2\omega+3\right)\phi}\left[\left(\omega+2\right)\varepsilon+\omega p\right]-\frac{\omega}{2}\left(\frac{\dot{\phi}}{\phi}\right)^{2}\nonumber \\
 & \hphantom{=}+2H\frac{\dot{\phi}}{\phi}+\frac{k}{a^{2}}+\frac{1}{2\left(2\omega+3\right)\phi}\left(\phi\frac{dV}{d\phi}-2V {\color{blue}{+16\pi\frac{\phi}{c}\frac{dc}{d\phi}\frac{\mathcal{L}_{m}}{c}}} \right).
 \label{eq:Hdot(epsilon,p,phi)}
\end{align}

Moreover, in the cosmological context, Eq. (\ref{eq:EOM-phi}) reduces
to:
\begin{equation}
\ddot{\phi}+3H\dot{\phi}=\frac{1}{2\omega+3}\left[8\pi\frac{1}{c}\left(\varepsilon-3p\right)-\phi\frac{dV}{d\phi}+2V-16\pi\frac{\phi}{c}\frac{dc}{d\phi}\frac{\mathcal{L}_{m}}{c}\right].
\label{eq:EOM-phi-Cosmology}
\end{equation}
This equation determines the dynamics of $\phi$. It depends on $H=H\left(x^{0}\right)$,
which means that (\ref{eq:1stFriedmannEq}), (\ref{eq:Hdot(epsilon,p,phi)})
and (\ref{eq:EOM-phi-Cosmology}) should be solved simultaneously.
The latter also depends on the matter content through the factor $\left(\varepsilon-3p\right)$,
which demands us to specify the nature of the perfect fluid at play
in the cosmological era under scrutiny. Finally, Eq. (\ref{eq:EOM-phi-Cosmology})
exhibits a dependence on the potential $V\left(\phi\right)$ and its
derivative w.r.t. the field $\phi$. Ultimately, we will have to say
what is the form that we expect for potential $\phi$ which, in turn,
is written to include the physical couplings $G$ and $c$. This challenge
will be addressed next: In the following section we perform the phase
space analysis of our model. With regards to dynamical systems approach
in cosmology, we refer the reader to the interesting papers \citep{Bahamonde2018,Chatzarakis2019,Odintsov2017,Odintsov2017AccelMultiFluid,Odintsov2018,Odintsov2019,Oikonomou2018,Oikonomou2019,Oikonomou2020}. 

\section{The dynamics of $\phi$\label{sec:Dynamics}}

The potential $V\left(\phi\right)$ is assumed to be analytical but
otherwise completely general. It will be taken as dominant over the
contribution of the term involving the matter Lagrangian; more specifically:
\begin{equation}
\left|16\pi\frac{\phi}{c}\frac{dc}{d\phi}\frac{\mathcal{L}_{m}}{c}\right|\ll\left|-\phi\frac{dV}{d\phi}+2V\right|.
\end{equation}
This condition may be satisfied either for $\left|\frac{\phi}{c}\frac{dc}{d\phi}\right|\ll1$
or $\left|\frac{\mathcal{L}_{m}}{c}\right|\ll\left|-\phi\frac{dV}{d\phi}+2V\right|$.
We shall consider the first terms of the expansion of $V\left(\phi\right)$
in a power series:
\begin{equation}
V\left(\phi\right)=V_{0}+V_{1}\phi+V_{2}\phi^{2},\label{eq:V(phi)}
\end{equation}
where $V_{i}$ ($i=0,1,2$) are the first three constant coefficients
in the truncated series. This type of expansion up to second order in the power series is customary of the potential description around a local minimum of the function. It is a usual practice in classical mechanics and quantum mechanics to approximate the potential in the vicinity of a local minimum for that of a harmonic oscillator---e.g. Ref. \cite{Griffiths2018}. The usefulness of a power series expansion of the potential up to its first-order terms  is clearly noticeable in the inflationary context---e.g., \cite{Piatella2018}---with the quadratic term being a useful example to illustrate the mechanism of a slowly-rolling scalar field \cite{Baumann2022}. The power series expansion of $V$ has also applications in classical field theories, such as in Ref. \cite{Braaten2018}. A power series expansion up to second-order can be found in Ref. \cite{Piatella2018}; the interpretation of the different terms that participate in a truncated series such as that in Eq.~(\ref{eq:V(phi)}) are discussed in \cite{Faraoni2004,PeterUzan2009}, for instance. The form (\ref{eq:V(phi)}) encodes a few cases of special interest:
\begin{enumerate}
\item Let $V_{1}=V_{2}=0$. This realizes a constant potential that could
be rescaled by setting $V_{0}=0$.
\item For $V_{0}=V_{2}=0$ and $V_{1}=2\Lambda$ one recovers the ordinary
cosmological constant ($\Lambda=\text{constant}$) already present
in GR. The enticing feature here is that $V\left(\phi\right)=2\Lambda\phi$
would exhibit all the fundamental couplings in gravity $\left\{ G,c,\Lambda\right\} $
when we substitute $\phi=c^{3}/G$. (Nevertheless, recall that $\Lambda$
would still be a genuine constant here since it defines $V_{1}$.)
\item If $V_{0}=V_{1}=0$ and $V_{2}=\frac{1}{2}m_{\phi}^{2}$, the potential
accounts for a massive term in $\phi$'s field equation. In fact,
in this case $V\left(\phi\right)=\frac{1}{2}m_{\phi}^{2}\phi$ and
$\frac{d^{2}V}{d\phi^{2}}=m_{\phi}^{2}$ thus corroborating the interpretation.
The delicate part is to interpret the role of this mass in $\phi=c^{3}/G$.
It could be conjectured that the source of $\phi$'s mass is either
a massive photon or a massive graviton.
\end{enumerate}
In view of the above features, we consider the $V\left(\phi\right)$
in Eq. (\ref{eq:V(phi)}) adequate enough to allow for a sufficiently
general dynamical analysis of our scalar-tensor theory of covarying
physical couplings in gravity.

Substitution of Eq. (\ref{eq:V(phi)}) into (\ref{eq:EOM-phi-Cosmology})
yields a form for the field equation of $\phi$ that does not depend
on $V_{2}$. This is not the final possible simplification to this
equation. The next step further is to consider periods in the cosmic
history when 
\begin{equation}
\left|\frac{1}{c}\left(\varepsilon-3p\right)\right|\ll\left|\phi\frac{dV}{d\phi}-2V\right|,\label{eq:matter_versus_potential}
\end{equation}
or, more specifically,

\begin{equation}
\left(\varepsilon-3p\right)=0.\label{eq:VacuumRadiationConstraint}
\end{equation}
The above constraints not only guarantees a great simplification of
Eq. (\ref{eq:EOM-phi-Cosmology}) but it is also very reasonable.
Indeed, Eq. (\ref{eq:VacuumRadiationConstraint}) is consistent with
a radiation filled universe\textemdash described by the equation of
state $p=\varepsilon/3$. By abiding to Eq. (\ref{eq:VacuumRadiationConstraint})
we assume that the dynamics of the field $\phi$ takes place in this
era.

In standard cosmology, the radiation era spans from the period after
inflation and reheating all the way until the matter dominated phase.
It includes major events such as electroweak unification, quark-hadron
transition, neutrino decoupling, and nucleosynthesis \citep{Ellis2012}.
It is a key time period in the cosmic thermal history during which
the dynamics of $\phi$ would have taken place.

In any case, the essential ingredient to the remainder of our analysis
is that the condition in Eq. (\ref{eq:matter_versus_potential}) is
satisfied during all the dynamics of $\phi$. This means that the
combination $\left(\phi\frac{dV}{d\phi}-2V\right)$ built from the
potential $V\left(\phi\right)$ should dominate the matter-energy
content entering the play via $\frac{1}{c}T_{\mu\nu}$. This configuration
could be easily realized within varying-$c$ frameworks during regimes
where $\left(c/c_{0}\right)\gg1$; for example, in the very early
universe, before the phase transition of the decaying $c\left(x^{0}\right)$
admitted by some VSL models \citep{Uzan2003}.

Plugging (\ref{eq:VacuumRadiationConstraint}) into (\ref{eq:EOM-phi-Cosmology})
results in:

\begin{equation}
\ddot{\phi}+3H\dot{\phi}-\left(\frac{V_{1}}{2\omega+3}\right)\phi-\left(\frac{2V_{0}}{2\omega+3}\right)=0.\label{eq:EOM-phi-Dynamics}
\end{equation}
A solution to this equation leads to $\phi=\phi\left(x^{0}\right)$,
which constrains the dynamics of $\left(c^{3}/G\right)$. The physical
solution to (\ref{eq:EOM-phi-Dynamics}) must avoid the point $\phi=0$
since the notion of a vanishing speed of light or of an infinitely
large gravitational coupling is meaningless. In order to realize this
physical requirement, we displace the origin of the scalar field $\phi$
by defining:
\begin{equation}
\varphi\equiv\phi+2\frac{V_{0}}{V_{1}},\label{eq:varphi}
\end{equation}
where we admit $V_{0}\neq0$ and $V_{1}$ finite. In terms of this
new field variable, Eq. (\ref{eq:EOM-phi-Dynamics}) reads:
\begin{equation}
\ddot{\varphi}+3H\dot{\varphi}-\frac{V_{1}}{2\omega+3}\varphi=0.\label{eq:EOM-varphi-Dynamics}
\end{equation}
The solution to this equation is not so easy because $H=H\left(x^{0}\right)$
and is given by Eq. (\ref{eq:1stFriedmannEq}).\footnote{Even after we solve Eq. (\ref{eq:EOM-varphi-Dynamics}) for $\phi=\phi\left(x^{0}\right)$
using $H=H\left(x^{0}\right)$, we should have to justify a particular
ansatz for $c=c\left(x^{0}\right)$ which would lead to $G\left(x^{0}\right)$.} 

\subsection{Elements of stability theory for dynamical systems\label{subsec:Stability-theory}}

We shall study Eq. (\ref{eq:EOM-varphi-Dynamics}) from the point
of view of dynamical systems. For this reason, it is worth briefly
reviewing the elements of standard phase space analysis most commonly
used in modern gravity. This is done in the following subsections,
whose discussions are based on the papers \citep{Bahamonde2018},
\citep{Odintsov2017AccelMultiFluid}, and references therein. Check
also Ref. \citep{Hirsch2012}.

\subsubsection{Linear stability theory\label{subsubsec:Linear-stability-theory}}

Let us suppose a physical system can be described by $n$ variables
$\mathbf{y}=y^{1},y^{2},\ldots,y^{n}$. Each of these variables are
real-valued functions, so that $\mathbf{y}\in M\subset\mathbb{R}^{n}$.
Now consider that the dynamics of this physical system can be described
by a set of $n$ differential equations of first order:
\begin{equation}
\dot{\mathbf{y}}=\mathbf{f}\left(\mathbf{y}\right),\label{eq:y_dot}
\end{equation}
where dot indicates derivative with respect to a parameter representing
the evolution of this system, say a time coordinate ($x^{0}$). Here,
$\mathbf{f}\left(\mathbf{y}\right)$ represents a set of $n$ analytical
functions (which can be nonlinear) of the variables $\mathbf{y}$.
Whenever there is no explicit dependence of $\mathbf{f}$ with respect
to $x^{0}$, the set of equations is called an \textit{autonomous system}.

If there exists a point $\mathbf{y}_{0}\in M$ such that $\mathbf{f}\left(\mathbf{y}_{0}\right)=0$,
then 
\begin{equation}
\left.\dot{\mathbf{y}}\right|_{\mathbf{y}=\mathbf{y}_{0}}=0.\label{eq:y_dot(y_0)}
\end{equation}
If there is a time $x_{\left(f\right)}^{0}$ where $\mathbf{y}\left(x_{\left(f\right)}^{0}\right)=\mathbf{y}_{0}$
, then the dynamics of the system ceases and $\mathbf{y}\left(x^{0}\right)=\mathbf{y}_{0}$
for any value of $x^{0}>x_{\left(f\right)}^{0}$. This point is called
a \textit{fixed point}, \textit{critical point} or \textit{equilibrium point}. 

Whenever a system has a fixed point, it is interesting to analyze
the trajectories in a neighbourhood $U\subset M$ of $\mathbf{y}_{0}$.
Roughly speaking, if the trajectories in $U$ diverge from the fixed
point, it is considered \textbf{unstable}; if the trajectories converge
to $\mathbf{y}_{0}$, then it is \textbf{stable}.

The simplest stability analysis that can be carried out for such system
consists in considering the equations in a small region $U$ where
$\mathbf{f}\left(\mathbf{y}\right)$ can be expanded in a Taylor series
about $\mathbf{y}_{0}$ and approximated to first order. The advantage
of this method is that it allows for an analytical solution for the
approximated system (the drawback is that the approximation by a Taylor
series is not always a good approximation for a nonlinear system).
In $U$, we have:
\begin{align}
\dot{\mathbf{y}} & \approx\mathbf{f}\left(\mathbf{y}_{0}\right)+\left(y^{i}-y_{0}^{i}\right)\cdot\frac{\partial\mathbf{f}}{\partial y^{i}}\left(\mathbf{y}_{0}\right)\nonumber \\
 & =\left[\left(\mathbf{y}-\mathbf{y}_{0}\right)\cdot\text{grad}\right]\mathbf{f}\left(\mathbf{y}_{0}\right).\label{eq:y_dot_Taylor}
\end{align}
A change of variable, $\mathbf{z}=\mathbf{y}-\mathbf{y}_{0}$, can
always be proposed so that the equilibrium point is the origin of
the coordinate system in the new variable:
\begin{equation}
\dot{\mathbf{z}}\approx\left(\mathbf{z}\cdot\text{grad}\right)\mathbf{f}\left(0\right),\label{eq:z_dot}
\end{equation}
or in a matrix form:
\begin{equation}
\dot{\mathbf{z}}\approx\mathbf{M}\cdot\mathbf{z},\label{eq:z_dot(M)}
\end{equation}
\begin{equation}
\left(\mathbf{M}\right)=\left(\begin{array}{cccc}
\frac{\partial f^{1}}{\partial z^{1}} & \frac{\partial f^{1}}{\partial z^{2}} & \dots & \frac{\partial f^{1}}{\partial z^{n}}\\
\frac{\partial f^{2}}{\partial z^{1}} & \frac{\partial f^{2}}{\partial z^{2}} & \dots & \frac{\partial f^{2}}{\partial z^{n}}\\
\vdots & \vdots & \ddots & \vdots\\
\frac{\partial f^{n}}{\partial z^{1}} & \frac{\partial f^{n}}{\partial z^{2}} & \dots & \frac{\partial f^{n}}{\partial z^{n}}
\end{array}\right)_{\mathbf{y}_{0}}.\label{eq:Jacobian_Matrix}
\end{equation}
$\mathbf{M}$ is called the \textit{Jacobian matrix} or \textit{stability
matrix}. As long as $\det\mathbf{M}\neq0$, we can look for a new
transformation of variables $\mathbf{z}\rightarrow\mathbf{Z}$ where
the transformed Jacobian matrix becomes diagonal: the elements of
the diagonal are actually the eigenvalues, $\lambda_{i}$, of the
Jacobian matrix. In this case, the set of differential equations will
have the following solution:
\begin{equation}
Z^{i}=Z_{0}^{i}e^{\lambda_{i}x^{0}}.\label{eq:Z}
\end{equation}
If the real part of at least one eigenvalue is positive, then the
trajectories will diverge from the fixed point and the system is \textbf{unstable};
if the real part of all eigenvalues are negative, then the trajectories
converge to the equilibrium point and the system is \textbf{stable}.
If the real part of one of the eigenvalues is zero, then the system
can be either stable or unstable\textemdash in this case, each system
has to be considered separately.

The eigenvalues of the Jacobian matrix are called \textit{Lyapunov coefficients}
and the stability analysis essentially consists in the study of their
real parts. Notice that the Jacobian matrix is composed of constant
coefficients once $\mathbf{f}$ is not explicitly $x^{0}$-dependent.
The stability that is characterized by the Lyapunov coefficients is
restricted to a region where the linear approximation for $\mathbf{f}$
is valid.

The above considerations shall be employed in Subsection \ref{subsec:Phase-space-Linear-Theory}.

When the components of the stability matrix are not constant or when
the linear approximation of the Taylor series is not a good approximation,
other methods have to be applied. For instance, a powerful method
to establish the stability of dynamical system when the linear stability
analysis fails is the Lyapunov's method presented in the next section.

\subsubsection{Lyapunov's method\label{subsubsec:Lyapunov's-method}\protect\footnote{Subsection \ref{subsubsec:Linear-stability-theory} follows closely the treatment given in Ref. \citep{Bahamonde2018}.}}

This method is more powerful and general than the one in Subsection \ref{subsec:Phase-space-Linear-Theory}
in the sense that it does not rely on linear stability: It has the
capability of investigating both local and global instability. In
addition, Lyapunov's method applies to non-hyperbolic points too,
something the linear stability theory fails to accomplish. A critical
point (fixed point) $\mathbf{y}=\mathbf{y}_{0}\in M\subset\mathbb{R}^{n}$
of the dynamical system (\ref{eq:y_dot}),
\begin{equation}
\dot{\mathbf{y}}=\mathbf{f}\left(\mathbf{y}\right),\label{eq:y_dot_Lyapunov}
\end{equation}
is classified as a hyperbolic point if all of the eigenvalues of the
stability matrix have non-zero real part. Otherwise, it is said to
be non-hyperbolic. 

Lyapunov's method for checking stability of a dynamical system consists
in finding a \textit{Lyapunov function} $V\left(\mathbf{y}\right)$,
such that:
\begin{description}
\item [{(i)}] $V\left(\mathbf{y}\right)$ is differentiable in a neighbourhood
$U$ of $\mathbf{y}_{0}$;
\item [{(ii)}] $V\left(\mathbf{y}\right)>V\left(\mathbf{y}_{0}\right)$;
\item [{(iii)}] $\dot{V}\le0$, $\forall\mathbf{y}\in U$.
\end{description}
There is no known systematic way of deriving the Lyapunov function
$V\left(\mathbf{y}\right)$. If a Lyapunov function exists, the \textit{Lyapunov
Stability Theorem} establishes that the critical point $\mathbf{y}_{0}$
is a \textbf{stable} fixed point if the requirement $\dot{V}\leq0$
is fulfilled. The critical point $\mathbf{y}_{0}$ is an \textbf{asymptotically
stable} fixed point if there is a Lyapunov function $V(\mathbf{y})$
satisfying the criterium $\dot{V}<0$. In addition, if $\lim_{\left\Vert \mathbf{y}\right\Vert \rightarrow\infty}V\left(\mathbf{y}\right)\rightarrow\infty$,
$\forall\mathbf{y}\in U$, then the critical point $\mathbf{y}_{0}$
is classified as \textit{globally stable} or \textit{globally asymptotically
stable}, respectively. See Ref. \citep{Bahamonde2018}.

Notice that requirement \textbf{(iii)} can be cast into the form
\begin{equation}
\dot{V}\left(y_{1},...,y_{n}\right)=\frac{\partial V}{\partial y_{1}}\dot{y}_{1}+\dots+\frac{\partial V}{\partial y_{n}}\dot{y}_{n}=\text{grad}V\cdot\mathbf{f}\left(\mathbf{y}\right)\leq0.\label{eq:V_dot}
\end{equation}
Eq. (\ref{eq:y_dot_Lyapunov}) was used in the last step. The relation
in Eq. (\ref{eq:V_dot}) will be important in Subsection \ref{subsec:Phase-space-Lyaponov-Method}
below.

\subsection{Phase-space analysis with $H\left(x^{0}\right)\approx H_{*}$ via
the linear stability theory \label{subsec:Phase-space-Linear-Theory}}

We avoid the difficulty mentioned below Eq. (\ref{eq:EOM-varphi-Dynamics})
by assuming, in this subsection, that the Hubble function is constant
during the time interval in which the dynamics of the field $\phi$
occurs. In other words,
\begin{equation}
H\left(x^{0}\right)\approx H_{*}=\text{constant}.\label{eq:Hconstant}
\end{equation}
It should be emphasized that this condition is not as restrictive
as it may seem. It does not mean that $H\left(t\right)$ is always
constant, but only during the time interval of $\varphi$'s evolution.
We are not demanding the universe to be stationary (with $\dot{a}=\text{constant}$);
we are essentially assuming that $H$ evolves slowly during the time
it takes $\varphi$ to reach an eventual attractor point. Again, this
is a first-order phase-transition scenario common to several models
of VSL \citep{Uzan2003}. 

In order to show that the picture described in the previous paragraph
is achievable, let us analyze the dynamics of $\varphi$ in the phase
space resorting to the linear stability theory \citep{Bahamonde2018}.
Let
\begin{equation}
p_{\varphi}\equiv\dot{\varphi}\label{eq:p-phi}
\end{equation}
be the momentum associated to $\varphi$. Then, Eqs. (\ref{eq:EOM-varphi-Dynamics})
and (\ref{eq:p-phi}) become the following coupled system of equations:
\begin{equation}
\begin{cases}
\dot{\varphi}=p_{\varphi}\\
\dot{p}_{\varphi}=-3H_{*}p_{\varphi}+\frac{V_{1}}{2\omega+3}\varphi
\end{cases}.\label{eq:AutonomousSystem}
\end{equation}
This pair of differential equations has the structure of an autonomous
system \citep{Hirsch2012}. The dynamical system (\ref{eq:AutonomousSystem})
is already linear; consequently, the linear stability theory is not
just an approximation around the critical point(s) but rather an exact
description, completely adequate to the case in hand. The result stemming
from this analysis should not suffer from any pathologies, such as
those mentioned e.g. in Ref. \citep{Odintsov2018} \textemdash see
also \citep{Goriely2000}. A quick inspection of (\ref{eq:AutonomousSystem})
revels the equilibrium point:
\begin{equation}
p_{\varphi,0}=\varphi_{0}=0.\label{eq:EquilibriumPoint}
\end{equation}

The next step is to analyze the stability of the fixed point according
to the Lyapunov criterium \citep{Hirsch2012}. The Lyapunov coefficients
$\lambda_{i}$ are the eigenvalues of the Jacobian matrix (stability
matrix) related to the autonomous system \citep{Bahamonde2018}. They
are found by solving the characteristic equation associated to this
matrix, and read:
\begin{equation}
\lambda_{\pm}=\frac{3H_{*}}{2}\left(-1\pm\sqrt{1+\frac{V_{1}}{3}\frac{1}{\left(1+\frac{2}{3}\omega\right)}\left(\frac{2}{3H_{*}}\right)^{2}}\right).\label{eq:lambda_pm}
\end{equation}

The autonomous system can be considered stable if the real part of
both eigenvalues satisfy \citep{Bahamonde2018}
\begin{equation}
\text{Re}\left[\lambda_{\pm}\right]\leqslant0.\label{eq:StabilityCriterium}
\end{equation}
The analysis of the direction fields in the phase space allows us
to double check if stability is attained (in an enlarged region surrounding
the equilibrium point \citep{Odintsov2017AccelMultiFluid}) and if
it is consistent with the Lyapunov criterium. Here we assume that
\begin{equation}
H_{*}>0,\label{eq:Hstar_positive}
\end{equation}
 i.e. our universe is expanding. 

We split our analysis to cover different possible ranges of values
for the parameters $V_{1}$ and $\omega$.

\subsubsection*{Case 1: Negative valued $V_{1}$}

Here we take $V_{1}<0$. There are two subcases to be analyzed according
to the values of the parameter $\omega$. These subcases are considered
separately next.\textbf{ }

\paragraph*{Subcase 1.1: Negative valued $\omega$ with $\omega<-\frac{3}{2}$}

\begin{figure}[h]
\begin{centering}
\includegraphics[scale=0.5]{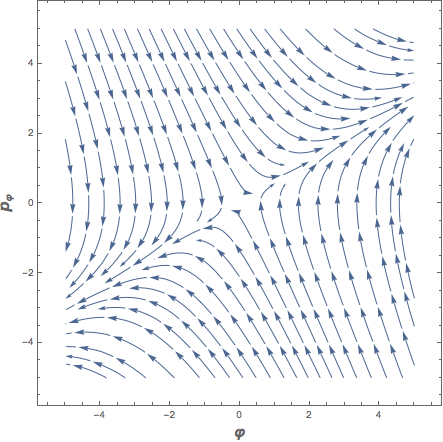}
\par\end{centering}
\caption{Direction fields in the phase space associated with the dynamics of
the field $\varphi$ for $V_{1}<0$ and $\omega<-\frac{3}{2}$. The
equilibrium point $\left(\varphi_{0},p_{\varphi,0}\right)=\left(0,0\right)$
is a repulsor of the trajectory lines. It is thus classified as an
unstable fixed point. }
\label{fig:VnWn}
\end{figure}

In the instance $V_{1}<0$ and $\omega<-\frac{3}{2}$, we may write:
\[
\frac{\left|V_{1}\right|}{3}\frac{1}{\left|\frac{2}{3}\left|\omega\right|-1\right|}>0.
\]
This condition is important to decide on the behaviour of the square-root
showing up in Eq. (\ref{eq:lambda_pm}). As a consequence,
\begin{equation}
\lambda_{+}=\frac{3H_{*}}{2}\left(-1+\sqrt{1+\frac{V_{1}}{3}\frac{1}{\left(1+\frac{2}{3}\omega\right)}\left(\frac{2}{3H_{*}}\right)^{2}}\right)>0.\label{eq:lambda_plus_VnWn}
\end{equation}
We conclude that $\lambda_{+}$ is a positive real number, so that
$\text{Re}\left[\lambda_{+}\right]>0$ is always satisfied, and the
associated equilibrium point is \textbf{unstable} according to Eq.
(\ref{eq:StabilityCriterium}). Since the analysis of $\lambda_{+}$
already reveals instability, there is no need to proceed to the study
of $\lambda_{-}$. The direction fields for the autonomous system
(\ref{eq:AutonomousSystem}) in the case where $V_{1}<0$ and $\omega<-\frac{3}{2}$
are shown in Fig. \ref{fig:VnWn}.

Fig. \ref{fig:VnWn} clearly shows that the trajectories diverge from
the equilibrium point $\left(\varphi_{0},p_{\varphi,0}\right)=\left(0,0\right)$,
thus indicating its instability in the context of the subcase under
scrutiny here. This conclusion will guide us towards constraining
the interval of values allowed for the parameters $V_{1}$ and $\omega$
if a reasonable physical interpretation for the dynamics of $\phi=c^{3}/G$
is to be established. We will decide on that later on, after we conclude
our stability analysis.

\paragraph*{Subcase 1.2: Values of $\omega$ satisfying $\omega>-\frac{3}{2}$}

Here we need both $V_{1}<0$ and $\omega>-\frac{3}{2}$. Under this
requirement, the eigenvalues in (\ref{eq:lambda_pm}) can be expressed
as:
\begin{equation}
\lambda_{\pm}=\frac{3H_{*}}{2}\left(-1\pm\sqrt{1-\frac{\left|V_{1}\right|}{3}\frac{1}{\left|1+\frac{2}{3}\omega\right|}\left(\frac{2}{3H_{*}}\right)^{2}}\right).\label{eq:lambda_pm_VnWp}
\end{equation}
The relevant part for deciding the nature (real or complex) of the
eigenvalues $\lambda_{\pm}$ is the second term within the square
root. If it is greater than 1, then
\[
\lambda_{\pm}=\frac{3H_{*}}{2}\left(-1\pm i\sqrt{\frac{\left|V_{1}\right|}{3}\frac{1}{\left|1+\frac{2}{3}\omega\right|}\left(\frac{2}{3H_{*}}\right)^{2}-1}\right).
\]
and both eigenvalues $\lambda_{\pm}$ become complex numbers with
negative real parts. Therefore, the equilibrium point is \textbf{stable},
cf. Eq. (\ref{eq:StabilityCriterium}). 

On the other hand, if
\begin{equation}
0<\frac{\left|V_{1}\right|}{3}\frac{1}{\left|1+\frac{2}{3}\omega\right|}\left(\frac{2}{3H_{*}}\right)^{2}\leqslant1,\label{eq:Term(V1,W,H)l1}
\end{equation}
the square root in (\ref{eq:lambda_pm_VnWp}) is a real number smaller
than one, and both eigenvalues $\lambda_{\pm}$ are real and negative.
This automatically satisfy the criterium $\text{Re}\left[\lambda_{\pm}\right]\leqslant0$
for stability, leading us to conclude, again, that the equilibrium
point $\left(\varphi_{0},p_{\varphi,0}\right)=\left(0,0\right)$,
is \textbf{stable}. 
\begin{figure}[h]
\begin{centering}
\includegraphics[scale=0.5]{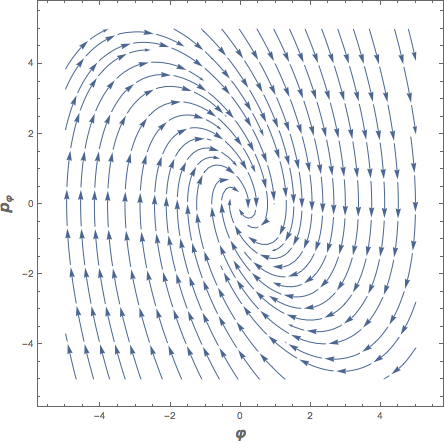}
\par\end{centering}
\caption{Direction fields representing the dynamics of the field $\varphi$
in the phase space for $V_{1}<0$ and $\omega>-\frac{3}{2}$. Trajectories
converge to $\left(\varphi_{0},p_{\varphi,0}\right)=\left(0,0\right)$
which is consequently a stable equilibrium point. }
\label{fig:VnWp}
\end{figure}

The direction fields in the plane $(\varphi,p_{\varphi})$ are analyzed
in Fig. \ref{fig:VnWp}. As we can see in the graph, the trajectories
in the phase space converge to the equilibrium point $\left(\varphi_{0},p_{\varphi,0}\right)=\left(0,0\right)$,
indicating its stability. The physical interpretation of this key
feature will be explored momentarily. Before that, we need to address
the case where $V_{1}$ is positive.

\subsubsection*{Case 2: Positive valued $V_{1}$}

The stability analysis of the equilibrium point $\left(\varphi_{0},p_{\varphi,0}\right)$
in the case $V_{1}>0$ also depends of the range of values assumed
by $\omega$. The two subcases are again $\omega<-\frac{3}{2}$ and
$\omega>-\frac{3}{2}$.

\paragraph*{Subcase 2.1: Negative valued $\omega$ with $\omega<-\frac{3}{2}$}

The case $\omega<-\frac{3}{2}$ yields:
\[
\lambda_{\pm}=\frac{3H_{*}}{2}\left(-1\pm\sqrt{1-\frac{V_{1}}{3}\frac{1}{\left|\frac{2}{3}\left|\omega\right|-1\right|}\left(\frac{2}{3H_{*}}\right)^{2}}\right),
\]
with $\frac{V_{1}}{3}\frac{1}{\left|\frac{2}{3}\left|\omega\right|-1\right|}>0$.
The analysis here follow exactly the same steps as for the subcase
1.2 above. If $\frac{V_{1}}{3}\frac{1}{\left|\frac{2}{3}\left|\omega\right|-1\right|}\left(\frac{2}{3H_{*}}\right)^{2}>1$,
the eigenvalues are complex numbers with negative real part, meaning
that the equilibrium point is \textbf{stable}. On the other hand,
if $\frac{V_{1}}{3}\frac{1}{\left|\frac{2}{3}\left|\omega\right|-1\right|}\left(\frac{2}{3H_{*}}\right)^{2}\leqslant1$,
the two eigenvalues are real negative numbers, implying that the system
has a \textbf{stable} fixed point.
\begin{figure}[h]
\begin{centering}
\includegraphics[scale=0.5]{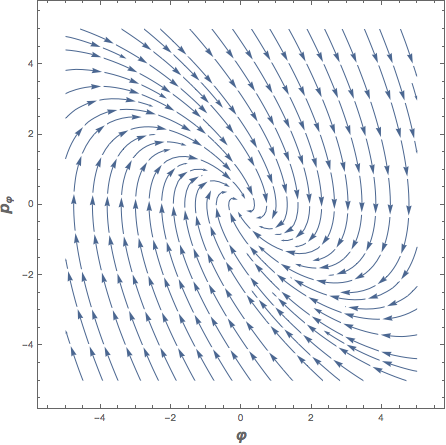}
\par\end{centering}
\caption{Direction fields in the phase space for $\phi$ in the case where
$V_{1}>0$ and $\omega<-\frac{3}{2}$. The point $\left(\varphi_{0},p_{\varphi,0}\right)=\left(0,0\right)$
is stable due to the convergence of trajectories. }
\label{fig:VpWn}
\end{figure}

The direction fields for the dynamics of $\varphi$ when $V_{1}>0$
and $\omega<-\frac{3}{2}$ are displayed in Fig. \ref{fig:VpWn}.
The trajectories converge to $\left(\varphi_{0},p_{\varphi,0}\right)=\left(0,0\right)$;
for this reason, it is classified as a stable equilibrium point.

\paragraph*{Subcase 2.2: Values of $\omega$ satisfying $\omega>-\frac{3}{2}$}

\begin{figure}[h]
\begin{centering}
\includegraphics[scale=0.5]{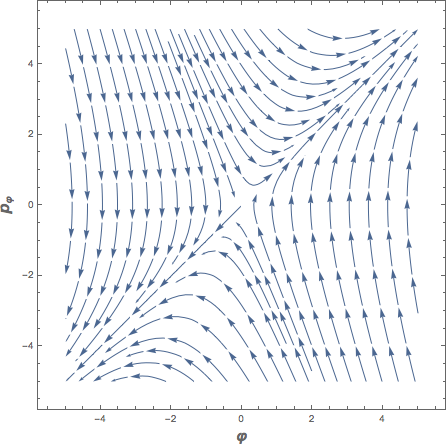}
\par\end{centering}
\caption{Direction fields in the phase space for $V_{1}>0$ and $\omega>-\frac{3}{2}$.
The direction field diverge from the point $\left(\varphi_{0},p_{\varphi,0}\right)=\left(0,0\right)$,
which is then classified as unstable.}
\label{fig:VpWp}
\end{figure}

In this situation, $V_{1}>0$ and $\omega>-\frac{3}{2}$, the eigenvalue
$\lambda_{+}$ is a real positive number. As a consequence, the stability
criterium $\text{Re}\left[\lambda_{\pm}\right]\leqslant0$ is violated
irrespective of the behaviour of the eigenvalue $\lambda_{-}$. Hence,
the equilibrium point $\left(\varphi_{0},p_{\varphi,0}\right)=\left(0,0\right)$
is \textbf{unstable}. This conclusion is confirmed by the direction
fields in Fig. \ref{fig:VpWp} showing trajectories diverging from
the equilibrium point. 

\subsection{Phase-space analysis with $H=H\left(x^{0}\right)$ via Lyapunov's
method \label{subsec:Phase-space-Lyaponov-Method}}

Presently, we generalize the autonomous system studied in Subsection
\ref{subsec:Phase-space-Linear-Theory} by allowing the Hubble function
to be $x^{0}$-dependent. In this case, the linear stability analysis
may be insufficient, or even misleading, as per Ref. \citep{Odintsov2018}
(and Ref. \citep{Bahamonde2018}). We are thus compelled to use other
methods of dynamical analysis. In particular, the Lyapunov's method\textemdash reviewed
in Subsection \ref{subsubsec:Lyapunov's-method}\textemdash proves to
be convenient in the present case. 

The generalization of the system in Eq. (\ref{eq:AutonomousSystem})
that takes into account $H=H\left(x^{0}\right)$ is:

\begin{equation}
\begin{cases}
\dot{\varphi}=p_{\varphi}\\
\dot{p}_{\varphi}=-3H\left(x^{0}\right)p_{\varphi}+\frac{V_{1}}{2\omega+3}\varphi
\end{cases}.\label{eq:DynamicalSystem}
\end{equation}
The equilibrium point is the same as in Eq. (\ref{eq:EquilibriumPoint}),
namely:
\begin{equation}
\mathbf{y}_{0}=\left(\begin{array}{c}
\varphi_{0}\\
p_{\varphi,0}
\end{array}\right)=\left(\begin{array}{c}
0\\
0
\end{array}\right).\label{eq:x_0}
\end{equation}
The (vector) function $\mathbf{f}\left(\mathbf{y}\right)$ that determines
the dynamics reads:
\begin{equation}
\mathbf{f}\left(\mathbf{y}\right)=\left(\begin{array}{c}
f_{1}\\
f_{2}
\end{array}\right)=\left(\begin{array}{cc}
0 & 1\\
\frac{V_{1}}{2\omega+3} & -3H\left(x^{0}\right)
\end{array}\right)\left(\begin{array}{c}
\varphi\\
p_{\varphi}
\end{array}\right),\label{eq:f(x)}
\end{equation}
where 
\begin{equation}
\mathbf{y}=\left(\begin{array}{c}
\varphi\\
p_{\varphi}
\end{array}\right),\label{eq:x}
\end{equation}
as it is immediately concluded from the comparison of the system (\ref{eq:DynamicalSystem})
with the definition (\ref{eq:y_dot_Lyapunov}).

Now we propose the following Lyapunov function $V\left(\mathbf{y}\right)$:

\begin{equation}
V\left(\mathbf{y}\right)=V\left(\varphi,p_{\varphi}\right)=a\varphi^{2}+bp_{\varphi}^{2},\label{eq:V(a,b)}
\end{equation}
with $a$ and $b$ constants. These constants will be determined in
accordance with the Lyapunov Stability Theorem momentarily.

The function in Eq. (\ref{eq:V(a,b)}) immediately satisfies the property
\textbf{(i)} in Subsection \ref{subsubsec:Lyapunov's-method}. Moreover,
notice that
\begin{equation}
V\left(\mathbf{y}_{0}\right)\rightarrow V\left(\varphi_{0},p_{\varphi0}\right)=V\left(0,0\right)=0.\label{eq:V(x_0)}
\end{equation}
This fact is relevant for checking if requirement \textbf{(ii)} is
fulfilled $\forall\mathbf{y}\in U$.

We can compute the first identity in Eq. (\ref{eq:V_dot}) for the
specific case of our dynamical system:
\begin{align}
\dot{V} & =\frac{\partial V}{\partial\varphi}\dot{\varphi}+\frac{\partial V}{\partial p_{\varphi}}\dot{p}_{\varphi}=2a\varphi\dot{\varphi}+2bp_{\varphi}\dot{p}_{\varphi}\nonumber \\
 & =2\left(a+\frac{V_{1}}{2\omega+3}b\right)p_{\varphi}\varphi-6bH\left(x^{0}\right)p_{\varphi}^{2},\label{eq:V_dot(a,b)}
\end{align}
where we have made use of Eq. (\ref{eq:DynamicalSystem}). The constant
$a$ and $b$ are arbitrary; we use this freedom to choose:
\begin{equation}
\left(a+\frac{V_{1}}{2\omega+3}b\right)=0,\qquad\text{and}\qquad b>0.\label{eq:Constraint(a,b)}
\end{equation}
Accordingly, Eq. (\ref{eq:V_dot(a,b)}) yields:
\begin{equation}
\dot{V}=-6bH\left(x^{0}\right)p_{\varphi}^{2}<0.\label{eq:V_dot_negative}
\end{equation}

The choice in (\ref{eq:Constraint(a,b)}) lead us to satisfy property
\textbf{(iii)} in Subsection \ref{subsubsec:Lyapunov's-method}. This
is so because the Hubble function is always positive for an expanding
Universe. Actually, we have $\dot{V}<0$ since $H\left(x^{0}\right)>0$
and $p_{\varphi}^{2}>0$, for $p_{\varphi}\neq0$. The equality in
\textbf{(iii)} would only be achieved if $b=0$, but this choice would
lead to $V=0$ in which case property \textbf{(ii)} in Subsection \ref{subsubsec:Lyapunov's-method}
would not be satisfied. 

\begin{figure}[h]
\includegraphics[scale=0.4]{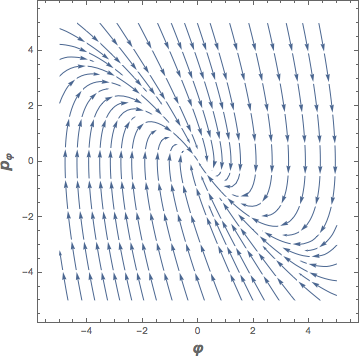}$\quad$\includegraphics[scale=0.4]{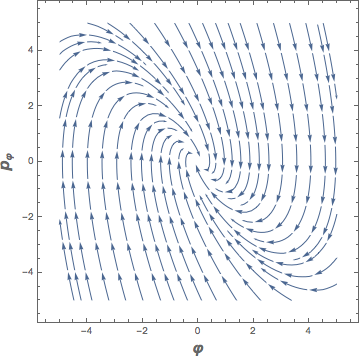}$\quad$\includegraphics[scale=0.4]{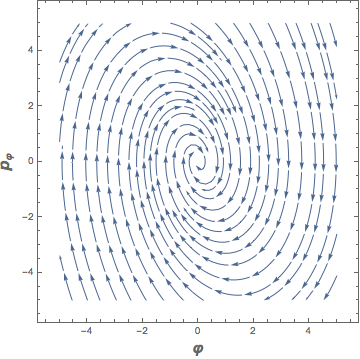}

\caption{Evolution of the direction fields toward the stable equilibrium point
at $\left(\varphi_{0},p_{\varphi,0}\right)=\left(0,0\right)$ in the
instance $V_{1}<0$ and $\omega>-\nicefrac{3}{2}$ for an $x^{0}$-dependent
Hubble parameter. The value of $H\left(x^{0}\right)$ decreases pregressively
in the plots from left to right. The direction fields in the sequence
of plots are consistent with the decreasing $H$: the damping effect
of the term containing $H$ in the field equation (\ref{eq:EOM-varphi-Dynamics})
of $\phi$ is reduced in the sequence of plots from left to right.
Consequently, the phase space diagrams look more and more alike that
of a (undamped) harmonic oscillator. }
\label{fig:Evolution_stability_H(x0)}
\end{figure}

Incidentally, it is the time to attempt to verify property \textbf{(ii)}.
Plugging Eq. (\ref{eq:Constraint(a,b)}) into (\ref{eq:V(a,b)}):
\begin{equation}
V\left(\varphi,p_{\varphi}\right)=\left(-\frac{V_{1}}{2\omega+3}\varphi^{2}+p_{\varphi}^{2}\right)b.\label{eq:V}
\end{equation}
Since $b>0$ and $\varphi^{2}>0$ and $p_{\varphi}^{2}>0$ in a neighbourhood
of the equilibrium point, the decision on the fulfilment of requirement
\textbf{(ii)} depends entirely on the values of parameters $V_{1}$
and $\omega$. Essentially, the stability will be achieved if
\begin{equation}
\frac{V_{1}}{2\omega+3}<0.\label{eq:Constraint(w,V1)}
\end{equation}
We identify four possibilities:
\begin{description}
\item [{(a)}] $V_{1}<0$ and $\omega<-\nicefrac{3}{2}$ then condition
(\ref{eq:Constraint(w,V1)}) is not satisfied, and the system is not
stable;
\item [{(b)}] $V_{1}<0$ and $\omega>-\nicefrac{3}{2}$ then condition
(\ref{eq:Constraint(w,V1)}) is satisfied, and the system is asymptotically
stable;
\item [{(c)}] $V_{1}>0$ and $\omega<-\nicefrac{3}{2}$ then the system
is asymptotically stable;
\item [{(d)}] $V_{1}>0$ and $\omega>-\nicefrac{3}{2}$ then the system
is not stable.
\end{description}
These possibilities are consistent with those unveiled by the linear
stability analysis performed in Subsection \ref{subsec:Phase-space-Linear-Theory}.
Here, however, the nature of the equilibrium point is kept even when
the Hubble parameter is allowed to change as a function of $x^{0}$.
Fig. \ref{fig:Evolution_stability_H(x0)} shows the direction fields
for different values of the Hubble function $H\left(x^{0}\right)$
and fixed values of $\omega$ and $V_{1}$ under condition \textbf{(b)}.

If we restrict our parameter space $\left(\omega,V_{1}\right)$ to
satisfy the conditions for stability, i.e. conditions \textbf{(b)}
or \textbf{(c)} above, then we verify that the equilibrium point $\mathbf{y}_{0}=\left(\varphi_{0},p_{\varphi,0}\right)=\left(0,0\right)$
is \textit{globally asymptotically stable}, since 
\[
\lim_{\sqrt{\varphi^{2}+p_{\varphi}^{2}}\rightarrow\infty}V\left(\varphi,p_{\varphi}\right)\rightarrow\infty,
\]
cf. the Lyapunov Stability Theorem\textemdash enunciated below point
\textbf{(iii)} in Subsection \ref{subsubsec:Lyapunov's-method}. 

\begin{figure}[h]
\begin{centering}
\textcolor{blue}{\includegraphics[scale=0.5]{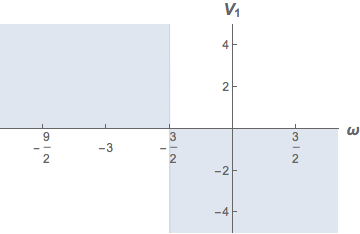}}
\par\end{centering}
\caption{Parameter space $\left(\omega,V_{1}\right)$ and the classification
of regions according to the stability of the dynamical system. The
shaded regions (in blue) highlight the values of the pair of parameters
$\left(\omega,V_{1}\right)$ rendering the fixed point $\left(\varphi_{0},p_{\varphi,0}\right)=\left(0,0\right)$
stable.}
\label{fig:Parameters_space_stability}
\end{figure}

Finally, we plot the parameter space of the pair $\left(\omega,V_{1}\right)$.
The stability of the system depends solely on the values of $V_{1}$
and $\omega$ both when $H\approx H_{*}$ (Subsection \ref{subsec:Phase-space-Linear-Theory})
and when $H=H\left(x^{0}\right)$ (this section). The shaded region
in Fig. \ref{fig:Parameters_space_stability} displays the region
of the parameter space where the dynamical system is stable.

\subsection{Discussion: the meaning of stability\label{subsec:Discussion}}

The results that have physical interest are those where the equilibrium
point is stable, i.e. the case where $V_{1}<0,\,\omega>-\frac{3}{2}$
and the case with $V_{1}>0,\,\omega<-\frac{3}{2}$ . These different
ranges of values for $V_{1}$ lead to two different interpretations.
Let us reconsider Eq. (\ref{eq:BD-likeFE}) and include the potential
(\ref{eq:V(phi)}) explicitly:
\begin{align}
R_{\mu\nu}-\frac{1}{2}g_{\mu\nu}R+\frac{V_{1}}{2}g_{\mu\nu} & =\frac{1}{c}\frac{8\pi}{\phi}T_{\mu\nu}-\frac{\left(V_{0}+V_{2}\phi^{2}\right)}{2}\frac{1}{\phi}g_{\mu\nu}\nonumber \\
 & +\frac{\omega}{\phi^{2}}\left(\nabla_{\mu}\phi\nabla_{\nu}\phi-\frac{1}{2}g_{\mu\nu}\nabla^{\rho}\phi\nabla_{\rho}\phi\right)\nonumber \\
 & +\frac{1}{\phi}\left(\nabla_{\mu}\nabla_{\nu}\phi-g_{\mu\nu}\square\phi\right).\label{eq:BD-likeFE(V)}
\end{align}

As mentioned in the beginning of Section \ref{sec:Dynamics}, we see
that the parameter $V_{1}$ plays the role of a cosmological constant
even in the dynamical phase of $\phi$. It may be suggestive to rename
this parameter $V_{1}\leftrightarrow2\Lambda$, in such a way that
the left-hand side of Eq. (\ref{eq:BD-likeFE(V)}) is formally the
same as the geometrical side of Einstein field equations. Hence, the
condition $V_{1}>0$ maps to $\Lambda>0$ which indicates that the
background solution to (\ref{eq:BD-likeFE(V)}) is a de Sitter-like
spacetime \citep{Weinberg1972}. Conversely, the condition $V_{1}<0$
would correspond to $\Lambda<0$ thus leading to AdS-like background.
(Of course, some conditions apply for achieving a strict dS/AdS background:
vacuum, derivatives of the field $\phi$ null, etc.).

In both cases, as long as we have stability of the equilibrium point,
the system will evolve so that the trajectories of field $\phi$ in
the phase space $\left(\varphi,p_{\varphi}\right)$ converge to $\left(\varphi_{0},p_{\varphi,0}\right)=\left(0,0\right)$.
When this fixed point is reached, we have simultaneously:
\begin{equation}
\begin{cases}
\dot{\phi}\rightarrow\dot{\phi}_{\text{eq}}=0\\
\phi\rightarrow\phi_{\text{eq}}=-2\frac{V_{0}}{V_{1}}=\text{constant}
\end{cases}.
\label{eq:eq_conditions}
\end{equation}
When the system gets to the equilibrium point, then
\begin{equation}
\frac{\dot{\phi}}{\phi}\rightarrow\frac{\dot{\phi}_{\text{eq}}}{\phi_{\text{eq}}}=-\left[\frac{\dot{G}}{G}-3\frac{\dot{c}}{c}\right]_{\text{eq}}=0,\label{eq:phi_dot_over_phi_eq}
\end{equation}
as implied from Eqs. (\ref{eq:eq_conditions}) and (\ref{eq:phi}).
This means that the condition
\begin{equation}
\frac{\dot{G}}{G}=3\frac{\dot{c}}{c}\label{eq:GuptasAnsatzVerified}
\end{equation}
will necessarily be satisfied whenever the system is in equilibrium:
Eq. (\ref{eq:GuptasAnsatz(G,c)}) holds true precisely for the parameter
value $\sigma=3$. This value is exact here: the uncertainty of $\sigma=3$
is theoretically zero, it is a natural consequence of the definition
$\phi$ demanded by basic dimensional analysis, and out of the dynamics
of $\phi$ in the phase space. 

Another consequence of this dynamics of $\phi$ is that the system
can start with arbitrary initial conditions. The condition $\frac{\dot{G}}{G}=3\frac{\dot{c}}{c}$
may not even be satisfied for this initial condition. It is just a
matter of time until the system evolves and converge to Eq. (\ref{eq:GuptasAnsatz(G,c)}).

We also point out the appearance of an effective cosmological constant
when the dynamics of $\phi$ relax to the equilibrium point (regardless
of the association $V_{1}\leftrightarrow2\Lambda$ suggested above).
In fact, the dynamics of the scalar field ceases at the equilibrium
point, i.e. $\nabla_{\mu}\phi=0$ with $\phi\rightarrow\phi_{\text{eq}}=\text{constant}$;
at this situation, the gravitational field equation (\ref{eq:BD-likeFE(V)})
approaches:
\begin{equation}
R_{\mu\nu}-\frac{1}{2}g_{\mu\nu}R+\left(\frac{V_{0}+V_{1}\phi_{\text{eq}}+V_{2}\phi_{\text{eq}}^{2}}{2\phi_{\text{eq}}}\right)g_{\mu\nu}=\frac{8\pi}{\phi_{\text{eq}}}T_{\mu\nu}^{\left(\text{eff}\right)},\label{eq:BD-likeFE_eq}
\end{equation}
where $T_{\mu\nu}^{\left(\text{eff}\right)}\equiv\frac{1}{c}T_{\mu\nu}$
is the conserved stress-energy tensor in our scalar-tensor model.
The term in the parenthesis above appears as an effective cosmological
constant,
\begin{equation}
\Lambda_{0}\equiv\left(\frac{V_{0}+V_{1}\phi_{\text{eq}}+V_{2}\phi_{\text{eq}}^{2}}{2\phi_{\text{eq}}}\right),\label{eq:Lambda0}
\end{equation}
and the field $\phi$, which converges to $\phi_{\text{eq}}$ in Eq.
(\ref{eq:eq_conditions}), can be set as 
\begin{equation}
\phi_{\text{eq}}=\frac{c_{0}^{3}}{G_{0}},\label{eq:phi_eq}
\end{equation}
and Eq. (\ref{eq:BD-likeFE_eq}) assumes the form of GR's field equation
with cosmological constant. The quantities $c_{0}$ and $G_{0}$ are
constant values of the couplings, which might as well be considered
as their present-day values in cosmological terms. In fact, due to
the last two equations, Eq. (\ref{eq:BD-likeFE_eq}) assumes the following
form at the present time:
\begin{equation}
R_{\mu\nu}-\frac{1}{2}g_{\mu\nu}R+\Lambda_{0}g_{\mu\nu}=\frac{8\pi G_{0}}{c_{0}^{4}}T_{\mu\nu},\label{eq:EFE-emergent}
\end{equation}
which is the familiar form of Einstein field equations, cf. Eq. (\ref{eq:EFE}). 

It is imperative that the following is cristal clear: the fact that
$\phi_{\text{eq}}=c_{0}^{3}/G_{0}=\text{constant}$ does not mean
that $G$ and $c$ will be constants after the equilibrium is reached.
All that is required thereafter is $\left(\dot{\phi}/\phi\right)=-\left(\dot{G}/G-3\dot{c}/c\right)=0$.
Accordingly, after the equilibrium is attained, it could be $c=c_{0}f$
and $G=G_{0}f^{3}$, where $f=f\left(x^{0}\right)$ is an arbitrary
function of the time coordinate: these time-dependent ansatze for
$G$ and $c$ satisfy both the requirements $\phi_{\text{eq}}=c_{0}^{3}/G_{0}$
and $\left(\dot{G}/G-3\dot{c}/c\right)=0$. 

In the face of the comments above, one concludes that, if an astrophysical
event takes place after the equilibrium condition is attained, it
is essentially impossible to identify the dynamics of $G$ separately
from the dynamics of $c$. In order to identify the eventual dynamics
of $G$ and $c$ with $\frac{\dot{G}}{G}\neq3\frac{\dot{c}}{c}$ (by
taking into account only gravitational fields equations, as we have
done here), one would have to consider situations out of the equilibrium
for $\phi$. If we are currently in the equilibrium condition, then
one would have to take into consideration events in the past history
of the universe. How far in that past one should go is still an open
question, hinging on the amount of time that the system takes to converge
to the equilibrium point. In any case, the condition in Eq. (\ref{eq:matter_versus_potential})
must be preserved for the dynamical analysis in Section \ref{sec:Dynamics}
to hold. As an example of what is stated above, we mention the solar
system tests constraining the Brans-Dicke parameter $\omega>40,000$
\citep{Will2018}, while as a dimensionless parameter, it would be
expected to be of order $\omega\approx1$. In these tests, the Parameterized
Post-Newtonian approximation is applied assuming that the dynamics
of the scalar field plays a role in the solar system evolution. In
our case, the dynamics ceased in the radiation era, way before the
solar system was formed. From this perspective, it would be meaningless
to try and discard our model based on BD constraints.

From the discussion above, the global picture of the dynamics in our
scalar-tensor model should be at sight. The system begins with arbitrary
initial conditions where $\phi=c^{3}/G$ is out of equilibrium. In
this phase, the functional form of $G$ and $c$ is not constrained
to any particular behaviour interweaved. There are two regions in the
parameter space of the pair $\left(\omega,V_{1}\right)$ enabling
the system to reach an stable critical point, cf. Fig. \ref{fig:Parameters_space_stability}.
By restricting our model parameters to these regions, one guarantees
that the stable equilibrium is reached, $\phi\rightarrow\phi_{\text{eq}}$,
and the couplings $G$ and $c$ are forced to evolve respecting $\left(\dot{G}/G-3\dot{c}/c\right)=0$
henceforth. The evolution of the universe from this point onward should
be impacted by the fact that $G$ scales as $c^{3}$. In the next
section we launch ourselves to the task of determining the background
cosmological evolution of our scalar-tensor model after the equilibrium.
Our final goal is to solve for the scale factor $a$ as a function
of the time-coordinate $x^{0}$. Along the way, we build the continuity
equation from the conservation of the effective stress-energy tensor
mentioned below Eq. (\ref{eq:BD-likeFE_eq})\textemdash see Subsection
\ref{subsec:Tmunu(eff)}. We will show that this continuity equation
differs from the conventionally used in FLRW cosmology for it contains
a term depending on $\dot{c}$. The ensuing VSL model is studied in
Subsection \ref{subsec:Co-Varying-VSL} assuming radiation and dust-matter
content. In both cases, the evolution of $a=a\left(x^{0}\right)$
tends to an accelerated de Sitter-type solution. As it happens, the
latter is true for two classes of VSL models: those with a decreasing
speed of light and those with an increasing $c=c\left(x^{0}\right)$.

\section{Analysis of the gravity field equation after $\phi$ reaches equilibrium\label{sec:Gravity-after-equilibrium}}

After $\phi$ reaches the equilibrium point, the field equation for
the gravitational part of our scalar-tensor model, Eq. (\ref{eq:BD-likeFE_eq}),
reduces to, i.e.

\begin{equation}
G_{\hphantom{\rho}\nu}^{\rho}=\frac{8\pi}{\phi_{\text{eq}}}\frac{1}{c}T_{\hphantom{\rho}\nu}^{\rho}-\Lambda_{0}\delta_{\hphantom{\rho}\nu}^{\rho},\label{eq:BD-likeFE-eq(Gmunu)}
\end{equation}
where $\phi_{\text{eq}}=\text{constant}$. However, $c=c\left(x^{0}\right)$
on the r.h.s. accompanying $T_{\mu\nu}$. In this section we investigate
non-vacuum cosmological solutions stemming from this new feature.

The 00-component and the $ii$-component of Eq. (\ref{eq:BD-likeFE-eq(Gmunu)}),
under the FLRW metric in (\ref{eq:FLRW}), read:
\begin{equation}
H^{2}=\frac{8\pi}{3\phi_{\text{eq}}}\frac{1}{c}\varepsilon+\frac{\Lambda_{0}}{3}-\frac{k}{a^{2}}\label{eq:Fried-1}
\end{equation}
and
\begin{equation}
\dot{H}+H^{2}=-\frac{4\pi}{3\phi_{\text{eq}}}\frac{1}{c}\left(\varepsilon+3p\right)+\frac{\Lambda_{0}}{3},\label{eq:Fried-2}
\end{equation}
which are the Friedmann equation and the acceleration equation, respectively.
(Recall that the definition $H=\dot{a}/a$ leads to $\ddot{a}/a=\dot{H}+H^{2}$.)

\subsection{The covariance of the effective stress-energy tensor\label{subsec:Tmunu(eff)}}

By taking covariant divergence of Eq. (\ref{eq:BD-likeFE-eq(Gmunu)})
and by using Bianchi identity $\left(\nabla_{\rho}G_{\hphantom{\rho}\nu}^{\rho}=0\right)$
and the metricity condition we conclude that:
\begin{equation}
\nabla_{\rho}\left(\frac{1}{c}T_{\hphantom{\rho}\nu}^{\rho}\right)=0.\label{eq:Tmunu(eff)-conservation}
\end{equation}
This is the extended conservation law in our scalar-tensor gravity
and justifies the definition of $T_{\mu\nu}^{\left(\text{eff}\right)}=\frac{1}{c}T_{\mu\nu}$
which is actually covariantly conserved. The effective stress-energy tensor $T_{\mu\nu}^{\left(\text{eff}\right)}$ is the Noether's current related to the symmetry of (infinitesimal) general coordinate transformations. When we specify Eq. (\ref{eq:Tmunu(eff)-conservation})
for the FLRW metric, we get:
\begin{equation}
\dot{\varepsilon}+3H\left(\varepsilon+p\right)=\left(\frac{\dot{c}}{c}\right)\varepsilon.\label{eq:ContinuityEq}
\end{equation}
The term on the r.h.s disappears for $\dot{c}=0$ so that Eq. (\ref{eq:ContinuityEq})
recovers the standard continuity equation of background cosmology
in GR in this particular case.

Eqs. (\ref{eq:Fried-1}), (\ref{eq:Fried-2}) and (\ref{eq:ContinuityEq})
form the essential set of equation of cosmology in our scalar-tensor
model allowing for the variation of $c$. Our proposal is to use the
first and the third equations in the set above to determine the cosmic
evolution. For that goal, two elements are necessary:
\begin{enumerate}
\item An equation of state $p=p\left(\varepsilon\right)$;
\item A constitutive equation for $c=c\left(x^{0}\right).$
\end{enumerate}
The EOS is taken as
\begin{equation}
p=w\varepsilon,\quad\text{where}\quad w=\begin{cases}
0, & \text{dust matter}\\
\frac{1}{3}, & \text{radiation}
\end{cases}.\label{eq:EOS}
\end{equation}
(Notice the distinction between $w$\textemdash the EOS parameter\textemdash and
$\omega$\textemdash the scalar-tensor constant parameter appearing
in Eq. (\ref{eq:S}) as the coefficient of the kinetic term of $\phi$.)
The continuity equation (\ref{eq:ContinuityEq}) then becomes:
\begin{equation}
\dot{\varepsilon}+\left[3H\left(1+w\right)-\left(\frac{\dot{c}}{c}\right)\right]\varepsilon=0.\label{eq:ContinuityEq(EOS)}
\end{equation}
The first term in the square brackets is the common contribution in
standard GR FLRW cosmology. The second term in the square brackets
is the contribution from our scalar-tensor model. The above equation
can not be solved explicitly without a constitutive equation for $c=c\left(x^{0}\right)$.
That is what we will do next.

\subsection{Modelling the varying speed of light\label{subsec:Co-Varying-VSL}}

In a few models realizing VSL scenarios, it is assumed that $c\left(x^{0}\right)$
suffered a first order phase transition in the begining of the thermal
history of the universe\textemdash see e.g., \citep{AlbrechtMagueijo1999}.
According to this image, the speed of light would have decreased its
magnitude as the scale factor increased. This motivates the proposal
in the next subsection. 

Before moving to the next subsection, it is probably worth emphasizing
that the assumption of a particular ansatz for $c=c\left(a\right)$,
which would be valid after the dynamics of the field $\phi$ ceases,
does not violate the equilibrium condition $\phi=\phi_{\text{eq}}={\rm constant}$.
For instance, if $c\sim1/a$ it suffices that $G\sim1/a^{3}$ for
maintaining $\phi=\frac{c^{3}}{G}={\rm constant}$.

\subsubsection{Speed of light scaling as the inverse of $a$ \label{subsec:c(inv_a)}}

Let us assume that
\begin{equation}
c=c_{0}\left(\frac{a_{0}}{a}\right).
\label{eq:c(inv_a)}
\end{equation}
That is: the speed of light decreases as the universe increases its
size. The consequence of (\ref{eq:c(inv_a)}) is \citep{Pipino2019}:
\begin{equation}
\frac{\dot{c}}{c}=-H.
\label{eq:c_dot_over_c(inv_a)}
\end{equation}
This is a reasonable choice that simplifies the continuity equation.
In fact, plugging (\ref{eq:c_dot_over_c(inv_a)}) into Eq. (\ref{eq:ContinuityEq(EOS)})
enables immediate integration to:
\begin{equation}
\varepsilon=\varepsilon_{0}\left(\frac{a_{0}}{a}\right)^{\left(4+3w\right)}.
\label{eq:epsilon(a)_c(inv_a)}
\end{equation}
In particular, $\varepsilon\sim a^{-5}$ in a radiation dominated
era ($w=1/3$). This violates the Stefan-Boltzmann law $\varepsilon\sim T^{4}$
because $a\sim T{}^{-1}$, with $T$ representing the temperature.
(Of course, the violation of the Stefan-Boltzmann law by radiation
in the model $c\sim a^{-1}$ occurs if we insist in a constant $k_{B}$.)
Curiously enough, dust-like matter ($w=0$) does recover the Stefan-Boltzmann
law $\varepsilon\sim a^{-4}$ in the our scalar-tensor model supplemented
by the ansatz $c\sim a^{-1}$. 

Substituting (\ref{eq:c(inv_a)}) and (\ref{eq:epsilon(a)_c(inv_a)})
into Eq. (\ref{eq:Fried-1}) with $k=0$ and $\phi_{\text{eq}}=\left(c_{0}^{3}/G_{0}\right)$
yields:
\begin{equation}
H^{2}=\frac{\Omega_{0}}{a_{0}^{2}}\left(\frac{a_{0}}{a}\right)^{3\left(1+w\right)}+\frac{\Lambda_{0}}{3}\label{eq:H(a)_c(inv_a)}
\end{equation}
under the definitions 
\begin{equation}
\varepsilon_{c,0}\equiv\frac{1}{a_{0}^{2}}\frac{3c_{0}^{4}}{8\pi G_{0}}\qquad\text{and}\qquad\Omega_{0}=\frac{\varepsilon_{0}}{\varepsilon_{c,0}},\label{eq:epsilon_c_and_Omega}
\end{equation}
where $\left[\varepsilon_{c,0}\right]=\left(\text{energy}\right)/\left(\text{length}\right)^{3}$
and $\Omega_{0}$ is dimensionless. Eq. (\ref{eq:H(a)_c(inv_a)})
integrates to
\begin{equation}
\left(\frac{a}{a_{0}}\right)\left[1+\sqrt{1+\frac{\Omega_{0}}{\left(a_{0}^{2}\frac{\Lambda_{0}}{3}\right)}\left(\frac{a_{0}}{a}\right)^{3\left(1+w\right)}}\right]^{2/3\left(1+w\right)}=\left[1+\sqrt{1+\frac{\Omega_{0}}{\left(a_{0}^{2}\frac{\Lambda_{0}}{3}\right)}}\right]^{2/3\left(1+w\right)}\exp\left[\sqrt{\frac{\Lambda_{0}}{3}}\left(x^{0}-x_{0}^{0}\right)\right].
\label{eq:a_c(inv_a)}
\end{equation}
In Eq. (\ref{eq:a_c(inv_a)}), $x_{0}^{0}$ is the value of the ``time''-coordinate
$x^{0}$ today. Eq. (\ref{eq:a_c(inv_a)}) is the exact solution for
the scale factor.

If $\Lambda_{0}a_{0}^{2}\gg\Omega_{0}$ (the cosmological constant
dominates over the matter-energy content), the above solution reduces
to:
\begin{equation}
a\simeq a_{0}\exp\left[\sqrt{\frac{\Lambda_{0}}{3}}\left(x^{0}-x_{0}^{0}\right)\right]\qquad\left(\Lambda_{0}a_{0}^{2}\gg\Omega_{0}\right).\label{eq:a_dS(inv_a)}
\end{equation}
This is a de Sitter-type accelerated expansion and could describe
dark energy. This is true regardless of the value of the parameter
$w$, which means that both a radiation era and a matter dominated
universe accommodate a de Sitter acceleration in the regime $\Lambda_{0}a_{0}^{2}\gg\Omega_{0}$.
Therefore, our scalar-tensor model with $c\sim a^{-1}$ is a promising
candidate for fitting observational data. It remains to be investigated
how the values of its parameters $\left(\Omega_{0},\Lambda_{0}\right)$
would be constrained by the data. Regardless, from the theoretical
point of view, one could imagine a cosmic evolution from a radiation-dominated
era to a matter-dominated universe to a dark-energy-dominated cosmos.
This last step would come as a natural evolution of the scale factor
in Eq. (\ref{eq:a_c(inv_a)}).

The question that poses itself at this point is the following. The
picture for the universe's evolution in the previous paragraph was
obtained from our scalar-tensor model and the particular ansatz $c\sim a^{-1}$
in which the speed of light decreases as the universe expands. How
dependent this scenario is on the particular choice of a decreasing
$c$? To put it another way: Could a different scenario with an increasing
speed of light accommodate radiation-dominated and matter-dominated
eras allowing for a subsequent dark-energy-type evolution?

In order to answer this question, we have to choose an ansatz for
$c=c\left(a\right)$ that is different from that in Eq. (\ref{eq:c(inv_a)}).
We will do this in the next section by assuming an ansatz introduced
by one of us (Gupta) in a series of papers \citep{Gupta2019,Gupta2022SNeIa,Gupta2022VCC,Gupta2020Cosmology,Gupta2021BBN,Gupta2022Quasars,Gupta2021Lensing,Gupta2021Orbital,Gupta2022FaintSun}.

\subsubsection{Speed of light scaling as the exponential of $a$ \label{subsec:c(exp_a)}}

Gupta's ansatz for an increasing speed of light has the form \citep{Gupta2020Cosmology}:
\begin{equation}
c=c_{0}\exp\left[\left(\frac{a}{a_{0}}\right)^{\alpha}-1\right],\label{eq:c(exp_a)}
\end{equation}
with $\alpha$ a positive constant of order one. Therefore,
\begin{equation}
\frac{\dot{c}}{c}=\alpha\left(\frac{a}{a_{0}}\right)^{\alpha}H.\label{eq:c_dot_over_c(exp_a)}
\end{equation}
Inserting (\ref{eq:c_dot_over_c(exp_a)}) into (\ref{eq:ContinuityEq(EOS)}):
\begin{equation}
\varepsilon=\varepsilon_{0}\left(\frac{a_{0}}{a}\right)^{3\left(1+w\right)}\left(\frac{c}{c_{0}}\right).\label{eq:epsilon(a)_c(exp_a)}
\end{equation}
For the standard non-varying speed of light picture, $c=c_{0}$, the
equation above recovers the traditional result from background FLRW
cosmology of GR. This means that the Stefan-Boltzmann law $\varepsilon\sim a^{-4}$
is recovered in the radiation era $\left(w=1/3\right)$ except for
the factor $\left(c/c_{0}\right)$. This is different from the case
in the previous subsection, where $c\sim a^{-1}$ and $\varepsilon\sim a^{-4}$
was realized only in the presence of dust-like matter.

Now, Eq. (\ref{eq:epsilon(a)_c(exp_a)}) can be used in the first
Friedmann equation (\ref{eq:Fried-1}) to yield the scale factor.
Indeed, taking $k=0$ and using the definitions of $\varepsilon_{c,0}$
and $\Omega_{0}$ in Eq. (\ref{eq:epsilon_c_and_Omega}) results in:
\begin{equation}
H^{2}=\frac{\Omega_{0}}{a_{0}^{2}}\left(\frac{a_{0}}{a}\right)^{3\left(1+w\right)}+\frac{\Lambda_{0}}{3}.\label{eq:H(a)_c(exp_a)}
\end{equation}
Surprisingly, this is exactly the same equation for $H$ derived in
the previous model\textemdash see Eq. (\ref{eq:H(a)_c(inv_a)}). Therefore,
the solution devised in Subsection \ref{subsec:c(inv_a)} for the model
with $c\sim a^{-1}$ is also valid in the present case, namely: Eq.
(\ref{eq:a_c(inv_a)}) leading to the same de Sitter-type limiting
case of Eq. (\ref{eq:a_dS(inv_a)}).

Now we are able to answer the question raised at the end of the last
section. A dark-energy phase is not a privilege of decreasing speed
of light scenarios. In our scalar-tensor model, an increasing speed
of light also accommodates an accelerated expansion. Therefore there
is enough freedom to imagine multiple eras of different kinematics
for $c=c\left(x^{0}\right)$, just like one admits different functions
of $a=a\left(x^{0}\right)$ for distinct eras of radiation or matter
domination. The universe might as well choose a decreasing speed of
light in the beginning of its evolution later evolving to a configuration
where $c$ increases as $x^{0}$ increases. 

By ``beginning of its evolution'' in the previous sentence we mean
the period immediately after $\phi$ reaches its equilibrium point
$\phi_{\text{eq}}$; after all, the whole formalism developed in the
present Section \ref{sec:Gravity-after-equilibrium} assumes the validity
of Eq. (\ref{eq:BD-likeFE-eq(Gmunu)}), which is the BD-like field
equation for the gravitational field after $\phi$ approaches $\phi_{\text{eq}}$.

It should be emphasized that the model in this subsection (Subsection
\ref{subsec:c(exp_a)}) was indeed tested against a plethora of astrophysical
and cosmological data with successes, including SNe Ia data \citep{Gupta2022SNeIa},
BAO and CMB data \citep{Gupta2020Cosmology}, BBN data \citep{Gupta2021BBN},
quasar data \citep{Gupta2022Quasars}, gravitational
lensing \citep{Gupta2021Lensing}, orbital mechanics \citep{Gupta2021Orbital},
the faint young-Sun problem \citep{Gupta2022FaintSun}.

\section{Concluding remarks\label{sec:Final-comments}}

In this paper, we studied a scalar-tensor model for gravity in which
both the gravitational coupling $G$ and the (causality) speed of
light $c$ are included in the scalar sector of the model through
the field $\phi=c^{3}/G$.

The field equations for $g_{\mu\nu}$ and $\phi$ are built and specified
for the homogeneous and isotropic cosmological background. The dynamics
of our field $\phi$ is then analyzed in the phase space under some
working hypotheses. We take the curvature parameter of the space sector
$k$ null. We use the first three terms in the series expansion of
the potential $V\left(\phi\right)$. We assume that the dynamics of
$\phi$ happens in a radiation era. We then study the evolution of
$\phi$ via linear stability theory (reviewed in Subsection \ref{subsec:Stability-theory})
assuming that the dynamical period of $\phi$ is short enough to assure
that Hubble function is approximately constant therein ({Subsection \ref{subsec:Phase-space-Linear-Theory}).
Finally, the dynamical system analysis is extended to allow for time-dependent
Hubble parameter; this requires the use of the Lyapunov's method (Subsection
\ref{subsec:Phase-space-Lyaponov-Method}). 

The phase space analysis shows that two sets of conditions upon the
free parameters of the model lead to trajectories converging to a
globally asymptotically stable equilibrium point. These conditions
are: (i) $V_{1}<0,\,\omega>-\frac{3}{2}$, and (ii) $V_{1}>0,\,\omega<-\frac{3}{2}$
, where $V_{1}$ is the constant coefficient of the linear term in
the $V\left(\phi\right)$ series and $\omega$ is the constant coefficient
of the kinetic term of $\phi$ in the model's action. The physical
interpretation of $V_{1}$ and $\omega$ are commented on momentarily. 

We have seen in Subsection \ref{subsec:Discussion} that $V_{1}$ could
be understood as a type of cosmological constant during the dynamical
phase of $\phi$; therefore, its positivity (or negativity) could
impact this version of $\Lambda$. Moreover, $V_{1}$ appears in the
effective cosmological constant $\Lambda_{0}$ which turns up at the
end of the dynamical evolution of $\phi$, when the trajectories of
the phase space converge toward the equilibrium point and stay there
for the rest of the cosmic history. The definition of $\Lambda_{0}$
includes the other coefficients of the potential too: $V_{0}$ is
the constant term in the $V\left(\phi\right)$ series, and $V_{2}$
is the coefficient of the quadratic term of $V$ in $\phi$. Our treatment
demands $V_{0}\neq0$ but otherwise unconstrained while $V_{2}$ is
totally unconstrained. The equilibrium value of the field, $\phi_{\text{eq}}=c_{0}^{3}/G_{0}$,
also enters the definition of $\Lambda_{0}$ which leads to the interesting
conclusion that the cosmological constant could depend on $\left(G_{0},c_{0}\right)$,
the (constant) equilibrium values of the gravitational coupling and
of the speed of light. 

The sign of the parameter $\omega$ is related to the physical nature
of the scalar field $\phi$. Indeed, if $\frac{\omega}{\phi}<0$ the
sign of the kinetic term for $\phi$ in the action indicates that
$\phi$ is a ghost field. This means that its Hamiltonian is unbounded
from below and it would be source of negative energy. The global sign
of the potential $V\left(\phi\right)$ is also important: together
with the sign of the kinetic term, it determines if $\phi$ could
also be an unphysical tachyon field breaking causality. In the vicinity
of the equilibrium point, $\phi>0$, so that $\omega<0$ would characterize
the field as a ghost field\textemdash see e.g., page 2 of \citep{Sbisa2015}
and references therein. This all means that we have to impose $\omega>0$
for a physically meaningful behaviour of $\phi$.

The big picture in our development is: after evolving to the equilibrium
stable point, the field $\phi$ reaches it equilibrium value $\phi_{\text{eq}}=c_{0}^{3}/G_{0}$
and stays there. Then, the dynamics of $\phi$ ceases and two things
happen: (i) the couplings $G$ and $c$ that are co-varying within
$\phi$ assume the dependence $G/G_{0}=\left(c/c_{0}\right)^{3}$,
where $G_{0}$ and $c_{0}$ could be interpreted as the numbers that
we use as fundamental constants today; and (ii) the gravitational
field equation of our scalar-tensor model, involving both $g_{\mu\nu}$
and $\phi$, degenerates to a field equation that is formally the
same as Einstein field equation of general relativity, but bears a
time-dependent factor $\left(1/c\right)$ alongside the ordinary energy
momentum tensor. At the present time $c=c_{0}$ and the field equation
recovers exactly Einstein equation of GR. That could explain why GR
is so successful in describing local and low-redshift data but should
be generalized in a larger scope of application. The relaxation of our
model toward a field equation that is formally the same as that of
GR is consistent with the works by Damour and Nordtvedt \citep{Damour1993PRL,Damour1993PRD}\textemdash see also Ref. \citep{Faraoni2022} by Faraoni and Franconnet.

When the dynamics of $\phi$ ends, $\dot{\phi}=0$, and the co-varying
$G$ and $c$ are forced to obey the relation $\left(\dot{G}/G\right)=\sigma\left(\dot{c}/c\right)$
with $\sigma=3$. This number is exact as far as the theoretical prediction
is concerned (under the hypotheses we have assumed, cf. the beginning
of Section \ref{sec:Dynamics}). Our conclusion is that at least some astrophysical phenomena would be unchanged if $G$ and $c$ vary concurrently while respecting
$\dot{\phi}=0$. This fact has been pointed out in Refs. \citep{Gupta2021Orbital,Gupta2021Lensing,Gupta2022VCC}, in relation with orbital timing, strong gravitational lensing of SNe Ia, and core-collapse supernova events. 

Our scalar-field model may be seen as a version of Brans-Dicke allowing
for a varying speed of light (alongside the varying $G$). Although
the field equations in Section \ref{sec:FieldEquations} look alike
those in BD theory, it must be underscored the fact that the varying
$c$ appears explicitly in the term containing the energy momentum
tensor. This forces the definition of an effective stress-energy tensor,
which is covariantly conserved thus assuring general covariance of
the gravity equation. This was discussed also in Section \ref{sec:Gravity-after-equilibrium},
where we build the Friedmann equations for the cosmology within our
scalar-tensor gravity. These equations are solved for two particular
ansatze of $c=c\left(x^{0}\right)$: (i) a speed of light that decreases
as $a$ increases, and (ii) a $c$ that increases as the universe
expands. It was shown that these both ansatze allow for an accelerated
expansion consistent with the effect of dark energy. The fact that
we were able to fully resolve the field equation for our scalar-tensor
gravity with varying coupling constants shows that the generality
of our model does not spoil practicality. The following logical step
in our research in the future will be to refine the details of the
picture by constraining the models with observational data. 

Future perspectives include to relax some of the hypotheses adopted
in this manuscript. One possibility is to let the dynamics of $\phi$
evolve in an era dominated by matter, for which $\left(\epsilon-3p\right)\neq0$
and the coupling of $c=c\left(x^{0}\right)$ with the matter Lagrangian
comes into effect. The eventual interrelation between the matter part
of the action and $c=c\left(x^{0}\right)$ brings forth the possibility
of $c$ being an independent scalar field. In this context, one would
have to take $G=G\left(x^{\mu}\right)$ and $c=c\left(x^{\mu}\right)$
as scalar fields evolving on their own (not enclosed within the field
$\phi$). To put it another way, in the stability analysis performed
in this paper we have one scalar degree of freedom, incarnated in
$\phi$; this is true even as both $G$ and $c$ are allowed to vary.
In a future paper, we will consider two independent degrees of freedom,
namely $G$ and $c$, with their own separate kinetic terms and potentials
(with or without a mutual interplay).

In this paper, the varying physical couplings were restricted to the
set $\left\{ G,c,\Lambda\right\} $\textemdash with the first two
forming the field $\phi$. Other works enlarge the set of possible
varying fundamental constants to $\left\{ G,c,\Lambda,\hbar,k_{B}\right\} $
and perform some phenomenological modelling in astrophysics and cosmology
\citep{Gupta2022VCC,Dannenberg2018}. Pursuing the fundamentals surrounding
an eventual variation of all these five couplings from a field theory
perspective is something that may be done in the future.

\bigskip{}
\noindent
{\textbf{Author Contributions:}} All authors have equally contributed to all parts of the paper. All authors have read and agreed to the published version of the manuscript.

\bigskip{}
\noindent
{\textbf{Funding:}} Macronix Research Corporation: 2020-23; National Council for Scientific and Technological Development–CNPq: 309984/2020-3.

\subsection*{Acknowledgments}

RRC thanks CNPq-Brazil (Grant 309984/2020-3) for partial financial
support. RRC and PJP are grateful to Prof. Rajendra P. Gupta and the
University of Ottawa for their hospitality. PJP acknowledges LAB-CCAM
at ITA. This work has been supported by a generous grant from Macronix
Research Corporation. The authors thank the valuable contributions of the anonymous referees whose comments and suggestions helped to improve the paper.


\section*{Appendix: Modified gravity field equations in cosmology}

Herein, we give additional steps in the derivations of Eqs.~(\ref{eq:1stFriedmannEq}), (\ref{eq:Hdot(epsilon,p,phi)}), and (\ref{eq:EOM-phi-Cosmology}) that are useful in the cosmological context from the general field equations (\ref{eq:BD-likeFE}), (\ref{eq:Tmunu}) and (\ref{eq:EOM-phi}) of our modified gravity with $\phi=c^3/G$---see Section \ref{sec:FieldEquations}.

\bigskip

Let us recall Eq. (\ref{eq:BD-likeFE}),
\begin{align}
R_{\mu\nu}-\frac{1}{2}g_{\mu\nu}R+\frac{V}{2\phi}g_{\mu\nu} & =\frac{1}{c}\frac{8\pi}{\phi}T_{\mu\nu}+\frac{\omega}{\phi^{2}}\left(\nabla_{\mu}\phi\nabla_{\nu}\phi-\frac{1}{2}g_{\mu\nu}\nabla^{\rho}\phi\nabla_{\rho}\phi\right)\nonumber \\
 & \hphantom{=\frac{1}{c}\frac{8\pi}{\phi}T_{\mu\nu}\;}+\frac{1}{\phi}\left(\nabla_{\mu}\nabla_{\nu}\phi-g_{\mu\nu}\square\phi\right),\label{eq:BD-likeFE-appendix}
\end{align}
and Eq. (\ref{eq:EOM-phi}),
\begin{equation}
\frac{2\omega}{\phi}\square\phi+R-\frac{\omega}{\phi^{2}}\nabla^{\rho}\phi\nabla_{\rho}\phi-\frac{dV}{d\phi}-16\pi\frac{1}{c^{2}}\frac{dc}{d\phi}\mathcal{L}_{m}=0.\label{eq:EOM-phi-appendix}
\end{equation}
We take the trace of Eq. (\ref{eq:BD-likeFE-appendix}) by contracting
with $g^{\mu\nu}$. Since $R=g^{\mu\nu}R_{\mu\nu}$ (and similarly
for $T_{\mu\nu}$) and $g^{\beta\mu}g_{\mu\nu}=\delta_{\hphantom{\beta}\nu}^{\beta}$,
it follows:
\begin{equation}
R=-\frac{1}{c}\frac{8\pi}{\phi}T+\frac{\omega}{\phi^{2}}\nabla^{\mu}\phi\nabla_{\mu}\phi+\frac{3\square\phi}{\phi}+\frac{2V}{\phi}.\label{eq:trace-BDFE}
\end{equation}
This can be substituted into Eq. (\ref{eq:EOM-phi-appendix}), yielding:
\begin{equation}
\square\phi=\frac{1}{2\omega+3}\left[\frac{1}{c}8\pi T+\phi\frac{dV}{d\phi}-2V+16\pi\frac{\phi}{c^{2}}\frac{dc}{d\phi}\mathcal{L}_{m}\right].\label{eq:phi-FE}
\end{equation} 

Spacetime homogeneity and isotropy requirements at the background
level are consistent with the FLRW metric, Eq. (\ref{eq:FLRW}),
\begin{equation}
ds^{2}=-\left(dx^{0}\right)^{2}+a^{2}\left(x^{0}\right)\left[\frac{dr^{2}}{1-kr^{2}}+r^{2}\left(d\theta^{2}+\sin^{2}\theta d\varphi^{2}\right)\right]\label{eq:FLRW-appendix}
\end{equation}
wherein the scale factor $a=a\left(x^{0}\right)$ is a function of
the ``time'' coordinate $x^{0}=ct$ alone. For the same reasons (spacetime
geometry symmetry properties), the scalar function $\phi$ will also
be a function exclusively of the $x^{0}$: $\phi=\phi\left(x^{0}\right)$.
The line element (\ref{eq:FLRW-appendix}) leads to \cite{Faraoni2004,Carroll2019,Baumann2022}:
\begin{equation}
\nabla^{\mu}\phi\nabla_{\mu}\phi=-\left(\frac{d\phi}{dx^{0}}\right)^{2}=-\dot{\phi}^{2},\label{eq:(nabla_phi)-square}
\end{equation}
\begin{equation}
\square\phi=-\left(\frac{d^{2}\phi}{dt^{2}}+3H\frac{d\phi}{dx^{0}}\right)=-\left(\ddot{\phi}+3H\dot{\phi}\right)=-\frac{1}{a^{3}}\frac{d}{dx^{0}}\left(a^{3}\dot{\phi}\right),\label{eq:Box_phi}
\end{equation}
and
\begin{equation}
R_{00}=-3\frac{\ddot{a}}{a}\,,\quad R_{11}=\frac{a\ddot{a}+2\dot{a}^{2}+2k}{1-kr^{2}},\quad R_{22}=r^{2}\left(a\ddot{a}+2\dot{a}^{2}+2k\right)=R_{33}/\sin^{2}\theta,\label{eq:Rmunu-FLRW}
\end{equation}
\begin{equation}
R=6\left(\dot{H}+2H^{2}+k/a^{2}\right),\label{eq:R-FLRW}
\end{equation}
with the Hubble function,
\begin{equation}
H=\frac{1}{a}\frac{da}{dx^{0}}=\frac{\dot{a}}{a},\label{eq:H-appendix}
\end{equation}
already defined in Eq. (\ref{eq:H}). Recall that an over dot denotes
derivative with respect to $x^{0}$ (and not simply a derivative with
respect to $t$). Moreover, we have stated around Eq. (\ref{eq:Tmunu-perfectfluid})
that
\begin{equation}
T_{\hphantom{\mu}\nu}^{\mu}=\text{diag}\left\{ -\varepsilon\left(x^{0}\right),p\left(x^{0}\right),p\left(x^{0}\right),p\left(x^{0}\right)\right\} ,\label{eq:Tmunu-perfectfluid-appendix}
\end{equation}
which trace reads
\begin{equation}
T=T_{\hphantom{\mu}\mu}^{\mu}=\left(-\varepsilon+3p\right).\label{eq:T}
\end{equation} 

With the above results it is straightforward to write the 00-component
of Eq. (\ref{eq:BD-likeFE-appendix}) in the form:
\begin{equation}
H^{2}=\frac{1}{c}\frac{8\pi}{3\phi}\varepsilon+\frac{\omega}{6}\left(\frac{\dot{\phi}}{\phi}\right)^{2}-H\frac{\dot{\phi}}{\phi}-\frac{k}{a^{2}}+\frac{V}{6\phi},\label{eq:1stFriedmannEq-appendix}
\end{equation}
which we have called the first Friedmann of our scalar-tensor cosmology
and corresponds to Eq. (\ref{eq:1stFriedmannEq}) in the main text. 

By making use of the result (\ref{eq:R-FLRW}) and Eq. (\ref{eq:trace-BDFE})
one gets:
\begin{equation}
\dot{H}+2H^{2}+\frac{k}{a^{2}}=-\frac{1}{c}\frac{4\pi}{3\phi}T-\frac{\omega}{6}\left(\frac{\dot{\phi}}{\phi}\right)^{2}+\frac{1}{2}\frac{\square\phi}{\phi}+\frac{V}{3\phi}.\label{eq:2ndFriedmannEq}
\end{equation}
Plugging Eqs. (\ref{eq:T}), (\ref{eq:phi-FE}), and (\ref{eq:1stFriedmannEq-appendix})
into (\ref{eq:2ndFriedmannEq}) gives:
\begin{align}
\dot{H} & =-\frac{1}{\left(2\omega+3\right)}\frac{1}{c}\frac{8\pi}{\phi}\left[\left(\omega+2\right)\varepsilon+\omega p\right]-\frac{\omega}{2}\left(\frac{\dot{\phi}}{\phi}\right)^{2}\nonumber \\
 & +2H\frac{\dot{\phi}}{\phi}+\frac{k}{a^{2}}+\frac{1}{2\left(2\omega+3\right)\phi}\left(\phi\frac{dV}{d\phi}-2V+16\pi\frac{\phi}{c^{2}}\frac{dc}{d\phi}\mathcal{L}_{m}\right).\label{eq:Hdot(epsilon,p,phi)-appendix}
\end{align}
This is Eq. (\ref{eq:Hdot(epsilon,p,phi)}) in the main part of the
paper. 

Finally, in the face of Eqs. (\ref{eq:Box_phi}) and (\ref{eq:T}),
Eq. (\ref{eq:phi-FE}) reduces to:
\[
-\left(\ddot{\phi}+3H\dot{\phi}\right)=\frac{1}{2\omega+3}\left[\frac{1}{c}8\pi\left(-\varepsilon+3p\right)+\phi\frac{dV}{d\phi}-2V+16\pi\frac{\phi}{c^{2}}\frac{dc}{d\phi}\mathcal{L}_{m}\right],
\]
i.e.
\begin{equation}
\ddot{\phi}+3H\dot{\phi}=\frac{1}{2\omega+3}\left[\frac{1}{c}8\pi\left(\varepsilon-3p\right)-\phi\frac{dV}{d\phi}+2V-16\pi\frac{\phi}{c^{2}}\frac{dc}{d\phi}\mathcal{L}_{m}\right],\label{eq:EOM-phi-Cosmology-appendix}
\end{equation}
thereby recovering Eq. (\ref{eq:EOM-phi-Cosmology}).

\bigskip


\end{document}